\documentstyle[prd,aps]{revtex}

\begin{document}

\title{The hypothesis of path integral duality II:~corrections
to quantum field theoretic results} 
\author{K.~Srinivasan\thanks{Electronic address:~srini@iucaa.ernet.in}, 
L.~Sriramkumar\thanks{Electronic address:~lsk@iucaa.ernet.in} and 
T.~Padmanabhan\thanks{Electronic address:~paddy@iucaa.ernet.in}}
\address{IUCAA, Post Bag 4, Ganeshkhind, Pune 411 007, INDIA.}

\maketitle

\begin{abstract}
In the path integral expression for a Feynman propagator of a spinless 
particle of mass~$m$, the path integral amplitude for a path of proper
length ${\cal R}(x,x'\vert g_{\mu\nu})$ connecting events~$x$ and~$x'$ 
in a spacetime described by the metric tensor~$g_{\mu\nu}$
is $\exp-\left[m\,{\cal R}(x,x'\vert g_{\mu\nu})\right]$.
In a recent paper, assuming the path integral amplitude to 
be invariant under the duality transformation ${\cal R} \to 
\left(L_P^2/{\cal R}\right)$, Padmanabhan has evaluated the 
modified Feynman propagator in an arbitrary curved spacetime.
He finds that the essential feature of this `principle of path 
integral duality' is that the Euclidean proper distance 
$(\Delta x)^2$ between two infinitesimally separated spacetime 
events is replaced by~$\left[(\Delta x)^2 + 4L_P^2\right]$. 
In other words, under the duality principle the spacetime behaves as 
though it has a `zero-point length'~$L_P$, a feature that is expected 
to arise in a quantum theory of gravity. 

In the Schwinger's proper time description of the Feynman propagator, 
the weightage factor for a path with a proper time~$s$ is~$\exp-(m^2s)$. 
Invoking Padmanabhan's `principle of path integral duality' 
corresponds to modifying the weightage factor $\exp-(m^2s)$ 
to~$\exp-\left[m^2s + (L_P^2/s)\right]$.
In this paper, we use this modified weightage factor in Schwinger's
proper time formalism to evaluate the quantum gravitational corrections 
to some of the standard quantum field theoretic results in flat and 
curved spacetimes.
In flat spacetime, we evaluate the corrections to:~(1)~the Casimir 
effect, (2)~the effective potential for a self-interacting scalar 
field theory, (3)~the effective Lagrangian for a constant electromagnetic
background and (4)~the thermal effects in the Rindler coordinates.
In an arbitrary curved spacetime, we evaluate the corrections 
to:~(1)~the effective Lagrangian for the gravitational field and 
(2)~the trace anomaly.
In all these cases, we first present the conventional result and then 
go on to evaluate the corrections with the modified weightage factor.
We find that the extra factor~$\exp-(L_P^2/s)$ acts as a regulator at
the Planck scale thereby `removing' the divergences that otherwise
appear in the theory. 
Finally, we discuss the wider implications of our analysis.
\end{abstract}

\pacs{PACS numbers:~}

\section{The principle of path integral duality}

All attempts to provide a quantum frame work for the gravitational
field have so far proved to be unsuccessful.
In the absence of a viable quantum theory of gravity, it is interesting
to ask whether we can say anything at all about the effects of metric 
fluctuations on quantum field theory in flat and curved spacetimes.

The Planck length$~L_P\equiv(G\hbar/c^3)^{1/2}$ is expected to play a 
vital role in the ultimate quantum theory of gravity.
(We shall set $\hbar=c=1$.)
Simple thought experiments clearly indicate that it is not possible 
to devise experimental procedures which will measure lengths with 
an accuracy greater than about~$O(L_P)$~\cite{paddy87}.  
Also, in some simple models of quantum gravity, when the spacetime 
interval is averaged over the metric fluctuations, it is found to be 
bounded from below at~$O(L_P^2)$~\cite{paddy85}.
These results suggest that one can think of the Planck length~$L_P$ 
as a `zero-point length' of spacetime.  
But how can such a fundamental length scale be introduced into
quantum field theory in an invariant manner?

The existence of a fundamental length implies that processes 
involving energies higher than Planck energies will be suppressed 
and the ultra-violet behavior of the theory will be improved.  
One direct consequence of such an improved behavior will be that 
the Feynman propagator will acquire a damping factor for energies 
higher than the Planck energy.  
The Feynman propagator $G_{\rm F}(x,x'\vert g_{\mu\nu})$ in a 
given background metric~$g_{\mu\nu}$ can be expressed in
Euclidean space as 
\begin{equation} 
G_{\rm F}(x,x'\vert g_{\mu\nu}) \equiv \sum_{\rm paths} 
\exp-\left[m\, {\cal R}(x,x'\vert g_{\mu\nu})\right],
\label{eqn:prop}
\end{equation}
where ${\cal R}(x,x'\vert g_{\mu\nu})$ is the proper length 
of a path connecting two events~$x$ and~$x'$.
The existence of a `zero-point length'~$L_P$ will correspond to the fact 
that paths with a proper length ${\cal R}\ll L_P$ will be suppressed 
in the path integral.
In two recent papers~\cite{paddyprl,paddyprd}, Padmanabhan has been able 
to capture exactly such a feature by assuming the weightage factor for a 
path of proper length~${\cal R}$ to be invariant under the duality
transformation~${\cal R} \to \left(L_P^2/{\cal R}\right)$.
On invoking such a `principle of path integral duality', the Feynman 
propagator above takes the following form:
\begin{equation}
G_{\rm F}(x,x'\vert g_{\mu\nu}) = \sum_{\rm paths} 
\exp-\left[m\,{\cal R}(x, x'\vert g_{\mu\nu})
+ \left({L_P^2}/{\cal R}(x,x'\vert g_{\mu\nu})\right)
\right].\label{eqn:modprop}
\end{equation}
(It can so happen that the fundamental length scale is not actually~$L_P$
but~$(\eta L_P)$, where $\eta$ is a factor of order unity.
Therefore, in this paper, when we say that the fundmental length is~$L_P$, 
we actually mean that it is of~$O(L_P)$.)
It turns out that the net effect of such a modification is to add the
quantity~$\left(-4L_P^2\right)$ to the Lorentzian proper 
distance~$(\Delta x)^2$ between two spacetime events $x$ and 
$x'$ in the original propagator.
In other words, the postulate of the path integral duality proves 
to be equivalent to introducing a `zero-point length' of~$O(L_P)$ 
in spacetime.

Two points needs to be stressed regarding the `principle of path 
integral duality'.
Firstly, this duality principle is able to introduce the fundamental 
length scale~$L_P$ into quantum field theory in an invariant manner.
Secondly, it is able to provide an ultra-violet regulator at the 
Planck scale thereby rendering the theory finite.
These two features of the duality principle can be illustrated with 
the example of the Feynman propagator in flat spacetime. 
In flat spacetime, the Feynman propagator~$G_{\rm F}(x,x'\vert 
\eta_{\mu\nu})$ can be expressed as a function of the Lorentz 
invariant spacetime interval~$(x-x')^2$. 
(The metric signature we shall use in this paper is such that 
$(x-x')^2>0$ for timelike events $x$ and $x'$.)
As we have mentioned in the last paragraph, invoking the `principle 
of path integral duality' corresponds to modifying the spacetime 
interval~$(x-x')^2$ to~$\left[(x-x')^2 -4L_P^2\right]$ in the 
Feynman propagator~\cite{paddyprl,paddyprd}.
Since~$\left[(x-x')^2 -4L_P^2\right]$ is again a Lorentz invariant
quantity, the modified Feynman propagator proves to be Lorentz invariant.
Also, in the limit of $x\to x'$ the original Feynman propagator has 
an ultra-violet divergence whereas the modified propagator stays finite.
Thus, the `principle of path integral duality' is not only
able to introduce the fundamental length scale~$L_P$ into quantum field 
theory thereby rendering it finite, it is able to do so in an invariant 
manner.

In Schwinger's proper time formalism, the Euclidean space 
Feynman propagator described by Eq.~(\ref{eqn:prop}) above 
can be expressed as follows~\cite{schwinger51}:
\begin{equation}
G_{\rm F}(x,x'\vert g_{\mu\nu})
=\int_{0}^{\infty}ds\, e^{-m^2 s}\, K(x,x';s \vert g_{\mu\nu}),
\label{eqn:gfnsch}
\end{equation}
where $K(x,x';s\vert g_{\mu\nu})$ is the probability amplitude 
for a particle to propagate from $x$ to $x'$ in a proper time 
interval~$s$ in a given background spacetime described by the 
metric tensor~$g_{\mu\nu}$. 
The path integral kernel $K(x,x';s\vert g_{\mu\nu})$ is given by
\begin{equation}
K(x,x';s \vert g_{\mu\nu}) 
\equiv \int {\cal D}x\; \exp-\left(\frac{m}{4} \int_{0}^{s}ds\, 
g_{\mu\nu}\, {\dot x}^{\mu}\, {\dot x}^{\nu}\right).\label{eqn:knl}
\end{equation}
In his paper~\cite{paddyprd}, Padmanabhan has shown that, in 
Schwinger's proper time formalism, invoking the `principle 
of path integral duality' corresponds to modifying the weightage 
given to a path of proper time~$s$ from $\exp-(m^2s)$ to 
$\exp-\left[m^2s + (L_P^2/s)\right]$.
That is, the modified Feynman propagator as defined 
in~Eq.~(\ref{eqn:modprop}) above is given by the following 
integral:
\begin{equation}
G_{\rm F}^{\rm P}(x,x'\vert g_{\mu\nu})
=\int_{0}^{\infty}ds\, e^{-m^2 s}\, e^{-(L_P^2K/s)}\;
K(x,x';s \vert g_{\mu\nu}),\label{eqn:modgfnsch}
\end{equation}
where $K(x,x';s \vert g_{\mu\nu})$ is the same as in~Eq.~(\ref{eqn:knl}).

Consider a $D$-dimensional spacetime described by the metric 
tensor~$g_{\mu\nu}$. 
Let us assume that a quantized scalar field $\Phi$ of mass $m$ 
satisfies the following equation of motion:
\begin{equation}
({\hat H}+m^2)\Phi=0,\label{eqn:ofmotion}
\end{equation}
where ${\hat H}$ is a differential operator in the $D$-dimensional 
spacetime.
Then, in Lorentzian space, an effective Lagrangian corresponding to 
the operator ${\hat H}$ can be defined as follows~\cite{schwinger51}:
\begin{equation}
{\cal L}_{\rm corr}=-\frac{i}{2}
\int_0^{\infty}\frac{ds}{s}\;e^{-im^2s}\;
K(x, x; s\vert g_{\mu\nu}),\label{eqn:lcorr}
\end{equation}
where
\begin{equation}
K(x, x; s \vert g_{\mu\nu}) \equiv \langle x\vert 
e^{-i{\hat H}s} \vert x\rangle.\label{eqn:kernel}
\end{equation}
(If the quantum scalar field interacts either with itself or with 
an external classical field then ${\cal L}_{\rm corr}$ will prove 
to be the correction to the Lagrangian describing the classical 
background.)
The quantity $K(x, x; s\vert g_{\mu\nu})$ is the path integral 
kernel (in the coincidence limit) of a quantum mechanical system 
described by time evolution operator ${\hat H}$.
The integration variable $s$ acts as the time parameter for the 
quantum mechanical system.

We saw above that, in Euclidean space, invoking the `principle of 
path integral duality' corresponds to modifying the weightage factor
$\exp-(m^2s)$ to $\exp-\left[m^2s+(L_P^2/s)\right]$.
In Lorentzian space, this modified weightage factor is given 
by:~$\exp-\left[im^2s -i (L_P^2/s)\right]$.
Therefore, invoking the duality principle corresponds to 
modifying the effective Lagrangian ${\cal L}_{\rm corr}$ 
given by Eq.~(\ref{eqn:lcorr}) above to the following form:
\begin{equation}
{\cal L}_{\rm corr}^{\rm P}=-\frac{i}{2}
\int_0^{\infty}\frac{ds}{s}\; e^{-im^2 s}\; 
e^{iL_P^2/s}\;K(x, x; s\vert g_{\mu\nu})\label{eqn:lcorrpl},
\end{equation}
where $K(x, x; s\vert g_{\mu\nu})$ is still given by Eq.~(\ref{eqn:kernel}).
Thus, we have been able to introduce the fundamental length scale~$L_P$
into standard quantum field theory.
Using this modified effective Lagrangian, we can evaluate the quantum 
gravitational corrections to standard quantum field theoretic results 
in flat and curved  spacetimes.

The following remarks are in order.
When we take into account the quantum nature of gravity, the metric 
fluctuates.
Therefore, evaluating the quantum gravitational corrections to 
the quantum field theoretic results would mean averaging over 
these metric fluctuations.
In some simple models of quantum gravity, it has been shown that 
averaging the spacetime interval~$(\Delta x)^2$ over the metric 
fluctuations corresponds to modifying this spacetime interval 
to~$\left[(\Delta x)^2-L_P^2\right]$~\cite{paddy85}.
As we have pointed out earlier, the `principle of path 
integral duality' is able to achieve exactly such a feature.
Also, in the last paragraph, we saw that invoking the duality 
principle corresponds to simply modifying the weightage factor 
in the effective Lagrangian.
Therefore, we say that the effective Lagrangian with the modified 
weightage factor as given by Eq.~(\ref{eqn:lcorrpl}) above contains 
the quantum gravitational corrections.

This paper is organized as follows.
In the following six sections, we evaluate the quantum gravitational
corrections to some of the quantum field theoretic results in flat and 
curved spacetimes.
In sections~\ref{sec:casimir}--\ref{sec:rindler}, we evaluate the 
corrections to: (1)~the Casimir effect, (2)~the effective potential for
a self-interacting scalar field theory, (3)~the effective Lagrangian for 
a constant electromagnetic background and (4)~the thermal effects in the 
Rindler coordinates, in flat spacetime.
In sections~\ref{sec:grav} and~\ref{sec:trace}, we evaluate the quantum 
gravitational corrections to: (1)~the gravitational Lagrangian and (2)~the 
trace anomaly in an arbitrary curved spacetime.
In all these sections, we shall first present the conventional result and 
then go on to evaluate the corrections with the modified weightage factor.
Finally, in section~\ref{sec:cnclsns}, we discuss the wider implications 
of our analysis.

\section{Corrections to the Casimir effect}\label{sec:casimir}

The presence of a pair of conducting plates alters the vacuum 
structure of the quantum field and as a result there arises a 
non-zero force of attraction between the two conducting plates.
This effect is called the Casimir effect.
In the conventional derivation of the Casimir effect, the difference 
between the energy in the Minkowski vacuum and the Casimir vacuum is 
evaluated and a derivative of this energy difference gives the force 
of attraction between the Casimir plates (see, for e.g., Ref.~\cite{iandz80},
pp.~138--141).
To evaluate the Casimir force in such a fashion we need to know the
normal modes of the quantum field with and without the plates.
Therefore, if we are to evaluate the quantum gravitational corrections to 
the Casimir effect by the method described above, then we need to know as 
to how metric fluctuations will modify the modes of the quantum field. 
But, we only know as to how the quantum gravitational corrections 
can be introduced in the effective Lagrangian. 
Therefore, in this section, we shall first present a derivation of the 
Casimir effect from the effective Lagrangian approach and then go on 
to evaluate the quantum gravitational corrections to this effect with 
the modified weightage factor.
The system we shall consider in this section is a massless scalar field 
in flat spacetime.
Also, we shall evaluate the effective Lagrangian for two cases:~(i)~with
vanishing boundary conditions on a pair of parallel plates and~(ii)~with 
periodic boundary conditions on the quantum field.

\subsection{Vanishing boundary conditions on parallel plates}

\subsubsection{Conventional result}

Consider a pair of plates situated at $z=0$ and $z=a$.
Let us assume that the scalar field~$\Phi$ vanishes on these plates.
For such a case, the operator~${\hat H}$ corresponds to that of a free 
particle with the condition that its eigen functions along the $z$-direction 
vanish at $z=0$ and $z=a$.
Along the other $(D-1)$~perpendicular directions the operator~${\hat H}$ 
corresponds to that of a free particle without any boundary conditions.
The complete quantum mechanical kernel can then be written as
\begin{equation}
K(x, x'; s)= K_z(z, z'; s)\times K_{\perp}(x_{\perp}, x_{\perp}'; s).
\label{eqn:kernelfull}
\end{equation}
(We shall refer to the flat spacetime kernel $K(x,x';s\vert \eta_{\mu\nu})$ 
simply as $K(x,x';s)$.) 
In the limit $x_{\perp}\to x_{\perp}'$, 
$K_{\perp}(x_{\perp}, x_{\perp}'; s)$ is given by
\begin{equation}
K_{\perp}(x_{\perp}, x_{\perp}; s)
=\left(\frac{i}{(4\pi is)^{(D-1)/2}}\right).\label{eqn:kernelperp}
\end{equation}
The quantum mechanical kernel $K_z(z, z'; s)$ along the $z$-direction 
corresponds to that of a particle of mass $(1/2)$ in an infinite square 
well potential with walls at $z=0$ and $z=a$.
For such a case, the normalized eigen functions for an energy 
eigen value $E=(n \pi^2/a^2)$ are then given by (see, for e.g., 
Ref.~\cite{qmtext}, p.~65)
\begin{equation}
\Psi_E(z)=\sqrt{\frac{2}{a}}\; \sin(n\pi z/a),
\quad{\rm where}\quad n=1,2,3\ldots. 
\end{equation}
The corresponding kernel can then be written using the Feynman-Kac 
formula as follows (see, for instance, Ref.~\cite{fandh65}, p.~88):
\begin{eqnarray}
K(z,z';s)&=&\sum_{E} \Psi_E(z)\; \Psi_E^*(z')\; e^{-iEs}\nonumber\\
&=&\left(\frac{2}{a}\right) \sum_{n=1}^{\infty}\sin(n\pi z/a)\;
\sin(n\pi z'/a)\; e^{-(in^2 \pi^2 s/a^2)}\nonumber\\
&=&\left(\frac{1}{2a}\right)
\sum_{n=-\infty}^{\infty} \left\{e^{in\pi (z-z')/a}
-e^{in\pi (z+z')/a} \right\}\; e^{-(in^2 \pi^2 s/a^2)}.
\end{eqnarray}
Using the Poisson sum formula (see, for instance, Ref.~\cite{mandf53}, 
Part~I, p.~483) this kernel can be rewritten as (see, 
for e.g., Ref.~\cite{schulman81}, p.~46)
\begin{eqnarray}
\lefteqn{K_z(z, z'; s)}\nonumber\\
& &\qquad\quad=\; \left(\frac{1}{(4\pi i s)^{1/2}}\right)\;
\sum_{n=-\infty}^{\infty}\biggl\{\exp \left[i(z-z'+2na)^2/4s\right] 
-\exp \left[i(z+z'+2na)^2/4s\right]\biggl\}
\end{eqnarray}
so that in the coincidence limit (i.e. as $z=z'$), this kernel 
reduces to
\begin{eqnarray}
K_z(z, z; s)
&=& \left(\frac{1}{(4\pi i s)^{1/2}}\right)\;
\sum_{n=-\infty}^{\infty}\biggl\{\exp(in^2 a^2/s) 
-\exp \left[i(z+na)^2/s\right]\biggl\}\nonumber\\
&=& \left(\frac{1}{(4\pi i s)^{1/2}}\right)\;
\left\{ 1+ 2 \sum_{n=1}^{\infty}  e^{(in^2a^2/s)} 
-\sum_{n=-\infty}^{\infty}e^{i(z+na)^2/s}\right\}.
\end{eqnarray}
Therefore, the complete kernel (in the coincidence limit) 
corresponding to the operator~${\hat H}$ with the conditions
that its eigen functions vanish at $z=0$ and $z=a$ is given by
\begin{eqnarray}
K(x, x; s)
&=& K_z(z, z; s)\times
K_{\perp}(x_{\perp}, x_{\perp}; s)\nonumber\\
&=& \left(\frac{i}{(4\pi is)^{D/2}}\right)\;
\left\{1+ 2 \sum_{n=1}^{\infty}  e^{(in^2a^2/s)} 
-\sum_{n=-\infty}^{\infty}e^{i(z+na)^2/s}\right\}.\label{eqn:kernelcas1}
\end{eqnarray}
Substituting this kernel in Eq.~(\ref{eqn:lcorr}) and setting $m=0$,
we obtain that
\begin{equation}
{\cal L}_{\rm corr}
= \left(\frac{1}{2 (4\pi i)^{D/2}}\right)
\int_{0}^{\infty} {ds \over s^{(D/2)+1}} 
\left\{1+ 2 \sum_{n=1}^{\infty}  e^{in^2a^2/s} 
-\sum_{n=-\infty}^{\infty}e^{i(z+na)^2/s}\right\}.
\end{equation}

In flat spacetime, in the absence of any boundary conditions on the 
quantum scalar field~$\Phi$, the quantum mechanical kernel for the 
operator~${\hat H}$ corresponds to that of a free particle. 
The kernel for such a case is given by
\begin{equation}
K^0(x, x; s) 
= \left(\frac{i}{(4\pi i s)^{D/2}}\right).\label{eqn:kernel0} 
\end{equation}
That is there exists a non-zero ${\cal L}_{\rm corr}$ even for a free 
field in flat spacetime.
Therefore, the flat spacetime contribution as given by
\begin{equation}
{\cal L}_{\rm corr}^0= \left(\frac{1}{2(4 \pi i)^{D/2}}\right)
\int_0^{\infty}\frac{ds}{s^{(D/2)+1}}\; 
e^{-im^2 s}\label{eqn:lcorr0}
\end{equation}
has to be subtracted from all ${\cal L}_{\rm corr}$.

On subtracting the quantity ${\cal L}_{\rm corr}^0$ (given by 
Eq.~(\ref{eqn:lcorr0}) above with $m$ set to zero) from the 
expression for ${\cal L}_{\rm corr}$, we obtain that
\begin{eqnarray}
{\bar {\cal L}}_{\rm corr}
&=&\left({\cal L}_{\rm corr} - {\cal L}_{\rm corr}^0\right)\nonumber\\
&=& \left(\frac{1}{2(4\pi i)^{D/2}}\right)
\int_{0}^{\infty} \frac{ds}{s^{(D/2)+1}} 
\left\{2 \sum_{n=1}^{\infty}  e^{(in^2a^2/s)} 
-\sum_{n=-\infty}^{\infty}e^{i(z+na)^2/s}\right\}\nonumber\\
&=&\left(\frac{1}{2(4\pi i)^{D/2}}\right)
\Biggl\{2 \sum_{n=1}^{\infty}  
\int_{0}^{\infty} \frac{ds}{s^{(D/2)+1}}\;e^{(in^2a^2/s)}\nonumber\\
& &\qquad\qquad\qquad\qquad\qquad\qquad\qquad 
-\;\sum_{n=-\infty}^{\infty}
\int_{0}^{\infty} \frac{ds}{s^{(D/2)+1}}\; e^{i(z+na)^2/s}\Biggl\}.
\end{eqnarray}
The integrals in the above result can be expressed in terms of the
Gamma function (see, for e.g., Ref.~\cite{gandr80}, p.~934).
Therefore
\begin{eqnarray}
{\bar {\cal L}}_{\rm corr}
&=&\left(\frac{\Gamma(D/2)}{2 (4\pi)^{D/2}}\right)
\left\{2 \sum_{n=1}^{\infty} (n^2a^2)^{-D/2} 
-\sum_{n=-\infty}^{\infty} (z+na)^{2.(-D/2)}\right\}\nonumber\\
&=&\left(\frac{\Gamma(D/2)}{2 (4\pi a^2)^{D/2}}\right)
\left\{2 \sum_{n=1}^{\infty} n^{-D} 
-\sum_{n=-\infty}^{\infty}
\left[n+(z/a)\right]^{-D}\right\}.\label{eqn:reglcorrcas1}
\end{eqnarray}

Now, consider the case $D=4$.
The first series in the above expression for 
${\bar {\cal L}}_{\rm corr}$ can be expressed 
in terms of the Riemann zeta function 
(see, for e.g., Ref.~\cite{arfken85}, p.~334) and, the 
second series can be summed using the following 
relation (cf.~Ref.~\cite{pbm86}, Vol.~I, p.~652):
\begin{equation}
\sum_{n=-\infty}^{\infty}(k+\alpha)^{-p}
= \left(\frac{\pi (-1)^{p-1}}{(p-1)!}\right)\;
\frac{d^{p-1}}{d\alpha^{p-1}}
\left[{\rm cot}(\pi\alpha)\right],\label{eqn:sum1}
\end{equation}
with the result 
\begin{equation}
{\bar {\cal L}}_{\rm corr}
=\left(\frac{\pi^2}{1440 a^4}\right)
-\left(\frac{\pi^2}{96 a^4\, {\rm sin}^4(\pi z/a)}\right)\;
\left(3-2\, {\rm sin}^2(\pi z/a)\right).
\end{equation}
 Let us now consider the case $D=2$.
For such a case, from Eqs.~(\ref{eqn:reglcorrcas1}) and~(\ref{eqn:sum1}),
we obtain that
\begin{equation}
{\bar {\cal L}}_{\rm corr}
=\left(\frac{\pi}{24 a^2}\right)
\left(1-3{\rm cosec}^2(\pi z/a)\right).
\end{equation}
Note that there for both the cases $D=4$ as well as $D=2$, there
arises a term that is dependent on $z$.
(Compare these results with Eqs.~(5.11) and (5.15) that appear
in~Ref.~\cite{fulling89}, p.~105.)

\subsubsection{Results with the modified weightage factor}

Let us now evaluate the effective Lagrangian with the modified weightage
factor.
Substituting the kernel~(\ref{eqn:kernelcas1}) in Eq.~(\ref{eqn:lcorrpl})
and setting $m=0$, we obtain that
\begin{equation}
{\cal L}_{\rm corr}^{\rm P}
= \left(\frac{1}{2 (4\pi i)^{D/2}}\right)
\int_{0}^{\infty} \frac{ds}{s^{(D/2)+1}}\; e^{iL_P^2/s} 
\left\{1+ 2 \sum_{n=1}^{\infty}  e^{in^2a^2/s} 
-\sum_{n=-\infty}^{\infty}e^{i(z+na)^2/s}\right\}.
\end{equation}

Just as in the case of~${\cal L}_{\rm corr}$, there exists a 
non-zero~${\cal L}_{\rm corr}^{\rm P}$ even for a free field 
in flat spacetime.
This flat spacetime contribution has to be subtracted 
from all ${\cal L}_{\rm corr}^{\rm P}$. 
It is given by
\begin{equation}
{\cal L}_{\rm corr}^{{\rm P}0}
= \left(\frac{1}{2 (4\pi i)^{D/2}}\right)
\int_0^{\infty} \frac{ds}{s^{(D/2)+1}}\;e^{-im^2 s}\; 
e^{iL_P^2/s}.\label{eqn:lcorrpl0}
\end{equation}
On subtracting this quantity (with~$m$ set to zero) from the expression
for~${\cal L}_{\rm corr}^{\rm P}$, we get
\begin{eqnarray}
{\bar {\cal L}}_{\rm corr}^{\rm P}
&=&\left({\cal L}_{\rm corr}^{\rm P} 
- {\cal L}_{\rm corr}^{{\rm P}0}\right)\nonumber\\
&=& \left(\frac{1}{2 (4 \pi i)^{D/2}}\right)
\int_{0}^{\infty} \frac{ds}{s^{(D/2)+1}}\; e^{iL_P^2/s} 
\left\{2 \sum_{n=1}^{\infty}  e^{in^2a^2/s} 
-\sum_{n=-\infty}^{\infty}e^{i(z+na)^2/s}\right\}\nonumber\\
&=& \left(\frac{1}{2 (4\pi i)^{D/2}}\right)
\Biggl\{2 \sum_{n=1}^{\infty}  
\int_{0}^{\infty} \frac{ds}{s^{(D/2)+1}}\;
e^{i\left[n^2a^2+L_P^2\right]/s}\nonumber\\ 
& &\qquad\qquad\qquad\qquad\qquad\qquad\qquad
-\;\sum_{n=-\infty}^{\infty}
\int_{0}^{\infty} \frac{ds}{s^{(D/2)+1}}\;
e^{i\left[(z+na)^2+L_P^2\right]/s}\Biggl\}.
\end{eqnarray}
As before, the integrals can be expressed in terms of the 
Gamma function with the result
\begin{eqnarray}
{\bar {\cal L}}_{\rm corr}^{\rm P}
&=&\left(\frac{\Gamma(D/2)}{2 (4 \pi)^{D/2}}\right)
\left\{2 \sum_{n=1}^{\infty} \left[n^2a^2+ L_P^2 \right]^{-D/2} 
-\sum_{n=-\infty}^{\infty} 
\left[(x+na)^2+L_P^2\right]^{-D/2}\right\}\nonumber\\
&=&\left(\frac{\Gamma(D/2)}{2(4\pi a^2)^{D/2}}\right)
\Biggl\{2 \sum_{n=1}^{\infty} \left[n^2+(L_P^2/a^2)\right]^{-D/2}\nonumber\\ 
& &\qquad\qquad\qquad\qquad\qquad\qquad\qquad
-\;\sum_{n=-\infty}^{\infty}
\left[\left(n+(z/a)\right)^2+(L_P^2/a^2)\right]^{-D/2}\Biggl\}.
\end{eqnarray}

The series in the above result cannot be written in closed form for 
the case $D=4$.
Consider the case $D=2$.
For such a case the series in ${\bar {\cal L}}_{\rm corr}^{\rm P}$
above can be expressed in a closed form.
Making use of the following two relations (cf.~Ref.~\cite{pbm86}, 
Vol.~1, p.~685):
\begin{equation}
\sum_{n=0}^{\infty}(n^2+\alpha^2)^{-1}
=\left(\frac{1}{2\alpha^2}\right) + \left(\frac{\pi}{2\alpha}\right) 
{\rm coth}(\pi\alpha)
\end{equation} 
and
\begin{equation}
\sum_{n=-\infty}^{\infty}\left[(n+\alpha)^2+\beta^2\right]^{-1}
=\left(\frac{\pi}{\beta}\right)\; {\rm sinh}(2\pi\beta)
\left[{\rm cosh}(2\pi\beta) -{\rm cos}(2\pi\alpha)\right]^{-1},
\end{equation}
we find that ${\bar {\cal L}}_{\rm corr}^{\rm P}$ can be expressed as
follows:
\begin{eqnarray}
{\bar {\cal L}}_{\rm corr}^{\rm P}
&=&\Biggl\{-\frac{1}{8 \pi L_P^2} 
+ \left(\frac{1}{8 a L_P}\right){\rm coth}(\pi L_P/a)\nonumber\\
& &\qquad\qquad\qquad\;\;
+\;\left(\frac{1}{8 a L_P}\right){\rm sinh}(2\pi L_P/a)
\left[{\rm cos}(2\pi z/a)-{\rm cosh}(2\pi L_P/a)\right]^{-1}\Biggl\}.
\end{eqnarray}
In the limit of $L_P\to 0$, we find that
\begin{eqnarray}
{\bar {\cal L}}_{\rm corr}^{\rm P}
\to & &\Biggl\{\left(\frac{\pi}{24 a^2}\right)
\left(1-3{\rm cosec}^2(\pi z/a)\right)\nonumber\\
& &\qquad\qquad\qquad\qquad
\;\; - L_P^2 \left(\frac{\pi^3}{360 a^4}\right)
\left(1+30{\rm cosec}^2(\pi z/a)
-45{\rm cosec}^4(\pi z/a)\right)\Biggl\}.
\end{eqnarray}
That is, the lowest order quantum gravitational corrections appear 
at $O\left(L_P^2/a^2\right)$.

\subsection{Periodic boundary conditions}

\subsubsection{Conventional result}

Let us now consider the case wherein we impose periodic boundary 
conditions on the quantum field along the $z$-direction.
That is, let us assume that the scalar field~$\Phi$ takes on the 
same value at the points $z$ and $(z+a)$. 
For such a case, the operator~${\hat H}$ corresponds to that of a 
free particle with the condition that its eigen functions along the 
$z$-direction take on the same value at $z$ and $(z+a)$.
Along the other $(D-1)$~perpendicular directions the operator~${\hat H}$ 
corresponds to that of a free particle without any boundary conditions.
The complete kernel can then be written as in Eq.~(\ref{eqn:kernelfull}).
where $K_{\perp}(x_{\perp},x_{\perp};s)$ is given by
Eq.~(\ref{eqn:kernelperp}).

The normalized eigen functions of the operator ${\hat H}$ along the 
$z$-direction corresponding to an energy eigen value $E=(4n^2 \pi^2 /a^2)$
are then given by
\begin{equation}
\Psi_E(z) = \sqrt{\frac{1}{a}}\, e^{(2i n \pi z/a)}
\quad{\rm where}\quad n=0, \pm 1, \pm 2, \ldots.
\end{equation}
The corresponding kernel can then be written using the Feynman-Kac 
formula as follows (see, for instance, Ref.~\cite{fandh65}, p.~88):
\begin{equation}
K(z,z';s) = \sum_{E} \Psi_E(z)\; \Psi_E^*(z')\; e^{-iEs}
=\left(\frac{1}{a}\right) \sum_{n=-\infty}^{\infty}
\exp \left[2i n \pi (z-z')/a\right]\; e^{-(4in^2 \pi^2 s/a^2)},
\end{equation}
which in the limit of $z\to z'$ reduces to
\begin{equation}
K(z,z;s)=\left(\frac{1}{a}\right) \sum_{n=-\infty}^{\infty}
e^{-(4in^2 \pi^2 s/a^2)}.
\end{equation}
Therefore, the complete kernel in $D$ dimensions is given by
\begin{equation}
K(x, x; s) = \left(\frac{i}{(4\pi i s)^{(D-1)/2}}\right)
\left(\frac{1}{a}\right)\; 
\sum_{n=-\infty}^{\infty} e^{-4in^2\pi^2 s/a^2}.
\end{equation}
Using the Poisson sum formula (see, for e.g., Ref.~\cite{mandf53}, 
Part~1, p.~483) the above sum can be rewritten as
\begin{equation}
K(x, x; s) = \left(\frac{i}{(4\pi i s)^{D/2}}\right) 
\sum_{n=-\infty}^{\infty} e^{in^2 a^2/4s}
=\left(\frac{i}{(4\pi i s)^{D/2}}\right) 
\left\{1+2\sum_{n=1}^{\infty} e^{in^2 a^2/4s}\right\}\label{eqn:kernelcas2}.
\end{equation}
Substituting this expression in Eq.~(\ref{eqn:lcorr}) and setting $m=0$, 
we obtain that
\begin{equation}
{\cal L}_{\rm corr}
=\left(\frac{1}{2(4\pi i)^{D/2}}\right) 
\int_0^{\infty} \frac{ds}{s^{D/2+1}} 
\left\{1 + 2 \sum_{n=1}^{\infty} e^{in^2 a^2/4s}\right\}.
\end{equation}
Now, subtracting the quantity ${\cal L}_{\rm corr}^0$ (given by 
Eq.~(\ref{eqn:lcorr0}) with $m=0$) from the above expression, 
we get  
\begin{equation}
{\bar {\cal L}}_{\rm corr}
=\left({\cal L}_{\rm corr} -{\cal L}_{\rm corr}^0\right)
=\left(\frac{1}{(4\pi i)^{D/2}}\right) 
\sum_{n=1}^{\infty} \int_0^{\infty} 
\frac{ds}{s^{D/2+1}}\; e^{in^2 a^2/4s}.
\end{equation}
The integral over $s$ can be expressed in terms of the Gamma function
(see, for instance, Ref.~\cite{gandr80}, p.~934),
so that
\begin{equation}
{\bar {\cal L}}_{\rm corr}
= \left(\frac{\Gamma(D/2)}{(\pi a^2)^{D/2}}\right)
\sum_{n=1}^{\infty} n^{-D}
=\left(\frac{\Gamma(D/2)}{(\pi a^2)^{D/2}}\right)\; \zeta(D),
\end{equation}
where $\zeta(D)$ is the Riemann zeta-function (see, for 
e.g., Ref.~\cite{arfken85}, p.~334).

Now, consider the case $D=4$.
For such a case, we find that
\begin{equation}
{\bar {\cal L}}_{\rm corr}
=\left(\frac{\Gamma(2)}{(\pi a^2)^{2}}\right)\; \zeta(4)
=\left(\frac{\pi^2}{90\, a^4}\right).
\end{equation}
Also, for the case $D=2$, ${\bar {\cal L}}_{\rm corr}$ is given by
\begin{equation}
{\bar {\cal L}}_{\rm corr}
=\left(\frac{\Gamma(1)}{(\pi a^2)}\right)\; \zeta(2)
=\left(\frac{\pi}{6 a^2}\right).
\end{equation}

\subsubsection{Results with the modified weightage factor}

Let us now evaluate the effective Lagrangian with the modified weightage
factor.
Substituting the kernel~(\ref{eqn:kernelcas2}) in~(\ref{eqn:lcorrpl}) and
setting $m=0$, we obtain that
\begin{equation}
{\cal L}_{\rm corr}^{\rm P}
=\left(\frac{1}{2(4\pi i)^{D/2}}\right) 
\int_0^{\infty} \frac{ds}{s^{D/2+1}}\;  e^{i L_P^2/s}\;
\left\{1 + 2 \sum_{n=1}^{\infty} e^{in^2 a^2/4s}\right\}.
\end{equation}
On subtracting the quantity ${\cal L}_{\rm corr}^{{\rm P}0}$ (given by
Eq.~(\ref{eqn:lcorrpl0}) with $m$ set to zero) from this expression,
we get
\begin{equation}
{\bar {\cal L}}_{\rm corr}^{\rm P}
= \left({\cal L}_{\rm corr}^{\rm P}-{\cal L}_{\rm corr}^{{\rm P}O}\right)
=\left(\frac{1}{(4\pi i)^{D/2}}\right) 
\sum_{n=1}^{\infty}
\int_0^{\infty} \frac{ds}{s^{D/2+1}}\;
e^{i(4 L_P^2+n^2 a^2)/4s}.
\end{equation}
As before, the integral over $s$ can be expressed in terms of the Gamma 
function with the result
\begin{equation}
{\bar {\cal L}}_{\rm corr}^{\rm P}
=\left(\frac{\Gamma(D/2)}{(\pi a^2)^{D/2}}\right)
\sum_{n=1}^{\infty}\left(n^2 + 4 L_P^2/a^2\right)^{-D/2}.
\end{equation}

Now, consider the case $D=4$.
For such a case
\begin{equation}
{\bar {\cal L}}_{\rm corr}^{\rm P}
=\left(\frac{1}{\pi^2 a^4}\right)
\sum_{n=1}^{\infty}\left(n^2+ 4 L_P^2/a^2\right)^{-2}.
\end{equation}
Using the following relation (cf.~Ref.~\cite{pbm86}, Vol.~I, p.~687):
\begin{eqnarray}
\sum_{n=0}^{\infty}\left(n^2+\alpha^2\right)^{-2}
&=& \left(\frac{1}{\alpha^4}\right) 
+ \sum_{n=1}^{\infty}(n^2+\alpha^2)^{-2}\nonumber\\
&=&\left(\frac{1}{2\alpha^4}\right)
+\left(\frac{\pi}{4\alpha^3}\right)\; {\rm coth}(\pi \alpha) 
+ \left(\frac{\pi^2}{4 \alpha^2}\right)\; {\rm cosech}^2(\pi \alpha),
\end{eqnarray}
we can express ${\bar {\cal L}}_{\rm corr}^P$ in a closed form as follows:
\begin{equation}
{\bar {\cal L}}_{\rm corr}^{\rm P}
=\left\{-\left(\frac{1}{32 \pi^2 L_P^4}\right)
+\left(\frac{1}{32 \pi a L_P^3}\right)\; {\rm coth}(2\pi L_P/a) 
+ \left(\frac{1}{16 a^2 L_P^2}\right)\; 
{\rm cosech}^2(2\pi L_P/a)\right\}.
\end{equation}
For the case $D=2$, ${\cal L}_{\rm corr}^{\rm P}$ is given by
\begin{equation}
{\bar {\cal L}}_{\rm corr}^{\rm P}
=\left(\frac{1}{\pi a^2}\right)
\sum_{n=1}^{\infty}\left(n^2+ 4 L_P^2/a^2\right)^{-1}.
\end{equation}
Using the following relation (cf.~Ref.~\cite{pbm86}. Vol.~I, p.~685)
\begin{equation}
\sum_{n=0}^{\infty}\left(n^2+\alpha^2\right)^{-1}
= \left(\frac{1}{\alpha^2}\right) 
+ \sum_{n=1}^{\infty}(n^2+\alpha^2)^{-1}
=\left(\frac{1}{2\alpha^2}\right)
+\left(\frac{\pi}{2\alpha}\right)\; {\rm coth}(\pi \alpha),
\label{eqn:relationone}
\end{equation}
we can express ${\bar {\cal L}}_{\rm corr}^{\rm P}$ in a closed 
form as follows:
\begin{equation}
{\bar {\cal L}}_{\rm corr}^{\rm P}
=\left\{-\left(\frac{1}{8 \pi L_P^2}\right)
+\left(\frac{1}{4 a L_P}\right)\; {\rm coth}(2\pi L_P/a)\right\}.
\end{equation}

Making use of the following series expansions (cf.~Ref.~\cite{gandr80}, 
p.~36)
\begin{equation}
{\rm coth}(\pi x) = \left(\frac{1}{\pi x}\right) 
+\left(\frac{2x}{\pi}\right) \sum_{n=1}^{\infty} (x^2 + n^2)^{-1}
\label{eqn:relationtwo}
\end{equation}
and
\begin{equation}
{\rm cosech}^2(\pi x) =\left(\frac{1}{\pi^2 x^2}\right)
+ \left(\frac{2}{\pi^2}\right)
\sum_{n-1}^{\infty}\left\{\frac{x^2 - n^2}{(x^2 + n^2)^{2}}\right\},
\label{eqn:relationthree}
\end{equation}
we find that, in the limit of $L_P\to 0$, 
${\bar {\cal L}}_{\rm corr}^{\rm P}$ reduces
to
\begin{equation}
{\bar {\cal L}}_{\rm corr}^{\rm P}
\to \left\{\left(\frac{\pi^2}{90 a^4}\right) 
-L_P^2 \left(\frac{8\pi^4}{945 a^6}\right)\right\}
\end{equation}
for the case $D=4$ while it reduces to
\begin{equation}
{\bar {\cal L}}_{\rm corr}^{\rm P}
\to \left\{\left(\frac{\pi}{6 a^2}\right) 
-L_P^2 \left(\frac{2\pi^3}{45 a^4}\right)\right\}
\end{equation}
for the case $D=2$. 
The lowest order quantum gravitational corrections to the conventional 
results appear at $O\left(L_P^2/a^2\right)$.

\section{The effective potential of a self-interacting scalar field
theory}\label{sec:phi4}

The system considered in this section is a massive, self-interacting 
scalar field in 4-dimensions (i.e. D=4) described by the action
\begin{eqnarray}
{\cal S}[\Phi] = \int d^4x\; {\cal L}(\Phi)
&=&\int d^4x \left\{\frac{1}{2}\partial^{\mu}\Phi \partial_{\mu}\Phi
-{\cal V}(\Phi)\right\}\nonumber\\
&=&\int d^4x \left\{\frac{1}{2}\partial^{\mu}\Phi \partial_{\mu}\Phi
-\frac{1}{2}m^2\Phi^2-{\cal V}_{int}(\Phi)\right\}
\end{eqnarray}
where $m$ is the mass and ${\cal V}_{int}(\Phi)$ represents the 
self-interaction of the scalar field.  
We are interested in studying the effects of small quantum fluctuations
present in the system around some classical solution $\Phi_c$. 
The classical solution will be assumed to be a constant or adiabatically 
varying so that its derivatives can be ignored.
The effect of these fluctuations will be studied by expanding the action
${\cal S}[\Phi]$ about the classical solution $\Phi_c$ and integrating over 
these fluctuations to obtain an effective potential.   
This effective potential will contain corrections to the original potential
${\cal V}(\Phi_c)$.      

Let $\Phi=(\Phi_c+\phi)$, where $\Phi_c$ is the classical solution.
The above action, when Taylor expanded about the classical solution, yields, 
\begin{eqnarray}
{\cal S}[\Phi]&\simeq& S[\Phi]\vert_{\Phi=\Phi_c} 
+ \left(\frac{\delta S[\Phi]}{\delta\Phi}\right)_{\Phi=\Phi_c}\!\!\!\!\phi
+ \frac{1}{2}\;
\left(\frac{\delta^2 S[\Phi]}{\delta\Phi^2}\right)_{\Phi=\Phi_c}\!\!\!\!
\phi^2\nonumber\\
&=&\int d^4x\Biggl\{\left(\frac{1}{2}\partial^{\mu}\Phi_c 
\partial_{\mu}\Phi_c -\frac{1}{2}m^2\Phi_c^2 
-{\cal V}_{int}(\Phi_c)\right)\nonumber\\
& &\qquad\qquad\qquad\qquad\qquad
+\; \left(\frac{1}{2}\partial^{\mu}\phi \partial_{\mu}\phi 
- \frac{1}{2}m^2\phi^2
-\frac{1}{2}{\cal V}_{int}''(\Phi_c)\,\phi^2\right)\Biggl\},
\label{eqn:siact}
\end{eqnarray}
where 
\begin{equation}
{\cal V}_{int}''(\Phi_c)\equiv
\left(\frac{\partial^2{\cal V}_{int}(\Phi)}{\partial 
\Phi^2}\right)_{\Phi=\Phi_c}
\end{equation}
and we have set $\left(\delta S[\Phi]/\delta\Phi\right)_{\Phi=\Phi_c}=0$, 
since $\Phi_c$ is the classical solution.
The correction to the classical action ${\cal S}[\Phi_c]$ is obtained
by integrating the degrees of freedom corresponding to the variable $\phi$.

Varying the action~(\ref{eqn:siact}) with respect to $\phi$ leads to the 
following equation of motion:
\begin{equation}
\left(\partial_{\mu}\partial^{\mu}+ m^2+{\cal V}_{int}''(\Phi_c)\right)\phi=0.
\end{equation}
From this equation the operator ${\hat H}$ can be easily identified to be
\begin{equation}
{\hat H}\equiv \left(\partial_{\mu}\partial^{\mu}
+{\cal V}_{int}''(\Phi_c)\right).
\end{equation} 
Since ${\cal V}_{int}''(\Phi_c)$ is a constant, the kernel corresponding 
to the operator ${\hat H}$ above can be immediately written down as
\begin{equation}
K(x, x; s)
=\left(\frac{1}{16\pi^2 is^2}\right)\;
\exp-\left[i {\cal V}_{int}''(\Phi_c)s\right].\label{eqn:potkernel}
\end{equation}
Using the above Kernel, the effective potential will be evaluated
by the conventional standard approach first and then by using the path 
integral duality principle.  

\subsection{Conventional result}

In the conventional approach, we substitute the kernel given in
Eqn.~(\ref{eqn:potkernel}) into Eqn.~(\ref{eqn:lcorr}) to obtain,
\begin{equation}
{\cal L}_{\rm corr}
=-\left(\frac{1}{32\pi^2}\right)
\int_{0}^{\infty}\frac{ds}{s^3}\;
e^{-i\alpha s}\qquad{\rm where}\qquad 
\alpha=\left(m^2+{\cal V}_{int}''(\Phi_c)\right).
\end{equation}
The above integral is quadratically divergent near the origin.  In the 
conventional approach, this integral is evaluated (after performing a
Euclidean rotation first) with the lower limit set to a small cutoff value 
$\Lambda$ with the limit $\Lambda \to 0$ considered subsequently. 
Only the leading terms that dominate in this limit
need to be retained.  Therefore, by performing repeated partial integrations,
we get,
\begin{equation}
{\cal L}_{\rm corr} = \left({1 \over 64\pi^2}\right) \left\{ \left[ 
{1 \over \Lambda^2} - {\alpha \over \Lambda} - \alpha^2 \ln(\Lambda \mu)
\right] - \alpha^2 \ln\left({\alpha \over \mu}\right) - \gamma \alpha^2 
\right\}
\end{equation}
where $\mu$ is an arbitrary finite parameter introduced to keep the argument
of the logarithms dimensionless and $\gamma$ is the Euler-Mascheroni constant. 
The last two terms within the curly brackets, namely, 
$\alpha^2\ln(\alpha /\mu)$ and $\gamma \alpha^2$, do not depend on 
$\Lambda$ and hence are finite in the limit $\Lambda \to 0$.
The term inside the square brackets however diverges. 
There are linear, quadratic and logarithmically divergent terms.
The quadratically divergent term is independent of $\alpha$ and 
being just an infinite constant can be dropped while the other two 
divergent terms depend on $\alpha$ and thus cannot be ignored. 

For an arbitrary ${\cal V}_{int}$, no sense can be made out of the above
expression for ${\cal L}_{\rm corr}$. 
One can only confine ourselves to those ${\cal V}_{int}$ for which the 
divergent quantities $\Lambda^{-1}\alpha$ and $\alpha^2 \ln(\Lambda \mu)$
have the same form as the original ${\cal V}(\Phi)$.  
If that is the case, the divergent terms can be absorbed into constants 
that determine the form of ${\cal V}(\Phi)$ and the theory can then be 
suitably reinterpreted. 
It is clear that for non-polynomial ${\cal V}(\Phi)$, the above criteria will
not be satisfied.  Therefore, assume that ${\cal V}(\Phi)$ is an $n$th-degree
polynomial in $\Phi$ with constants $\lambda_1, \lambda_2,\ldots ,\lambda_n$.
\begin{equation}
{\cal V}(\Phi) = \sum_{k=1}^{n} \lambda_k \Phi^k.
\end{equation}  
Then $\alpha = {\cal V}''$ will be a $(n-2)$ degree polynomial and 
$\alpha^2$ will be a $2(n-2)$ polynomial.  If the expression for 
${\cal L}_{\rm corr}$ is not to have terms originally not present in 
${\cal V}(\Phi)$, we must have 
\begin{equation}
2(n-2) \leq n 
\end{equation}
implying $ n\leq 4$.
Therefore, only if ${\cal V}(\Phi)$ is a polynomial of quartic degree 
or less can the divergences be absorbed and the theory reinterpreted.  
As an example, consider the case, 
\begin{equation}
{\cal V}_{int}(\Phi)=\frac{1}{4!}\lambda \Phi^4.
\end{equation}
Then
\begin{equation}
{\cal V}_{int}''(\Phi_c)= \frac{1}{2}\lambda \Phi_c^2\qquad
{\rm and}\qquad \alpha=\left(m^2+\frac{1}{2}\lambda \Phi_c^2\right).
\end{equation}
Carrying out the analysis, we get,
\begin{eqnarray}
{\cal V}_{\rm eff} &=& {\cal V}(\Phi_c) - {\cal L}_{\rm corr} \nonumber \\
&=& \frac{1}{2}m^2 \Phi_c^2 + \frac{1}{4!}\lambda \Phi_c^4 - 
{\cal L}_{\rm corr} \nonumber \\
&=& \frac{1}{2}m_{\rm corr}^2 \Phi_c^2 + \frac{1}{4!} \lambda_{\rm corr} 
\Phi_c^4 + {\cal V}_{\rm finite}
\end{eqnarray}
where 
\begin{eqnarray}
m_{\rm corr}^2 &=& m^2 + \frac{ \lambda}{32\pi^2} \left(\frac{1}{2\Lambda}
+ m^2 \ln(\Lambda \mu) \right) \nonumber \\
\lambda_{\rm corr} &=& \lambda + \frac{3 \lambda^2}{32\pi^2}\ln(\Lambda \mu)
\nonumber \\
{\cal V}_{\rm finite} &=& \frac{1}{64 \pi^2}\left( m^2 + \frac{\lambda}{2}
\Phi_c^2 \right)^2 \left\{\ln\left[ \frac{1}{\mu}\left( m^2 + \frac{\lambda}{2}
\Phi_c^2 \right)\right] + \gamma\right\} \label{eqn:corr}
\end{eqnarray}
The original potential ${\cal V}(\Phi)$ had two constants $m$ and $\lambda$,
which were the coefficients of $\Phi^2$ and $\Phi^4$. 
In ${\cal V}_{\rm eff}(\Phi_c)$ these are replaced by two other constants
$m_{\rm corr}$ and $\lambda_{\rm corr}$ which are functions of $m$, $\lambda$
and an arbitrary finite parameter $\mu$. 
These constants also contain divergent terms involving $\Lambda$.
Using renormalization group techniques, one can interpret ${\cal V}(\Phi)$
suitably. 
We will not discuss these techniques in this paper.

\subsection{Results with the modified weightage factor}

With the modified weightage factor, substituting the kernel given in 
Eq.~(\ref{eqn:potkernel}) in Eq.~(\ref{eqn:lcorrpl}), we obtain that
\begin{equation}
{\cal L}_{\rm corr}^{\rm P}
=-\left(\frac{1}{32\pi^2}\right)
\int_{0}^{\infty}\frac{ds}{s^3}\; e^{iL_P^2/s}\;
e^{-i\alpha s}\qquad{\rm where}\qquad 
\alpha=\left(m^2+{\cal V}_{int}''(\Phi_c)\right).
\end{equation}
The above integral can be easily performed and the resulting
${\cal L}_{\rm corr}^{\rm P}$ can be expressed in a closed form 
as follows (see, for instance, Ref.~\cite{gandr80}, p.~340):
\begin{equation}
{\cal L}_{\rm corr}^{\rm P}
=\left(\frac{\alpha}{16\pi^2 {L_P}^2}\right)\; 
K_2\left(2L_P\sqrt{\alpha}\right),
\end{equation}
where $K_2(2L_P\sqrt{\alpha})$ is the modified Bessel function.
Since 
\begin{equation}
{\cal L}_{\rm corr}^{{\rm P}0}
=-\left(\frac{1}{32\pi^2}\right)
\int_{0}^{\infty}\frac{ds}{s^3}\; e^{-im^2 s}\; e^{iL_P^2/s}
=\left(\frac{m^2}{16 \pi^2 L_P^2}\right)\;
K_2\left(2L_Pm\right),
\end{equation}
on subtracting this quantity from ${\cal L}_{\rm corr}^{\rm P}$, we
obtain that
\begin{eqnarray}
{\bar {\cal L}}_{\rm corr}^{\rm P}
&=&\left({\cal L}_{\rm corr}^{\rm P}
-{\cal L}_{\rm corr}^{{\rm P}0}\right)\nonumber\\
&=&\left(\frac{1}{16 \pi^2 L_P^2}\right)\;
\left\{\alpha\; K_2\left(2L_P\sqrt{\alpha}\right) 
- m^2\; K_2\left(2L_Pm\right)\right\}.
\end{eqnarray}
Since
\begin{eqnarray}
{\cal L}_{\rm eff}^{\rm P}
&=&\Biggl\{\frac{1}{2}\partial^{\mu}\Phi_c 
\partial_{\mu}\Phi_c-{\cal V}_{\rm eff}^{\rm P}\Biggl\}\nonumber\\
&=&\Biggl\{\left(\frac{1}{2}\partial^{\mu}\Phi_c 
\partial_{\mu}\Phi_c
-\frac{1}{2} m^2\Phi_c^2-{\cal V}_{int}(\Phi_c)\right)
+{\bar {\cal L}}_{\rm corr}^{\rm P}\Biggl\},
\end{eqnarray}
we have
\begin{equation}
{\cal V}_{\rm eff}^{\rm P}
=\Biggl\{\left(\frac{1}{2} m^2\Phi_c^2
+{\cal V}_{int}(\Phi_c)\right)
- \left(\frac{1}{16\pi^2 L_P^2}\right)\; 
\left(\alpha\; K_2\left(2L_P\sqrt{\alpha}\right)
-m^2\;K_2(2L_Pm)\right)\Biggl\}.
\end{equation}
The expression for the effective potential given above is applicable 
for arbitrary ${\cal V}_{int}$.  
This is in contrast to the conventional approach where the divergences 
appearing in ${\cal L}_{\rm corr}$ forced the potential to be a polynomial 
of quartic degree or less. 
Further, there is no need to introduce an arbitrary parameter $\mu$ as 
was required in the conventional approach.  
The need for such a parameter arose because of the cutoff $\Lambda$ that
was introduced by hand in order to isolate the divergent terms appearing
in ${\cal L}_{\rm corr}$.
The introduction of a fundamental length $L_P$ in the theory dispenses 
with such a need.  

Specialize now to the case when
\begin{equation}
{\cal V}_{int}(\Phi)=\frac{1}{4!}\lambda \Phi^4.
\end{equation}
Then
\begin{equation}
{\cal V}_{int}''(\Phi_c)= \frac{1}{2}\lambda \Phi_c^2\qquad
{\rm and}\qquad \alpha=\left(m^2+\frac{1}{2}\lambda \Phi_c^2\right).
\end{equation}  
For such a case, the corrections to the parameters of the theory can be
obtained from ${\cal V}_{\rm eff}^{\rm P}$ as follows:
\begin{equation}
{\left(m_{\rm corr}^{\rm P}\right)\!}^{2}
=\left(\frac{\partial^2 {\cal V}_{\rm eff}^{\rm P}}
{\partial\Phi_c^2}\right)_{\Phi_c=0}
=\left\{m^2-\left(\frac{\lambda}{16\pi^2 L_P^2}\right)\; K_2(2L_Pm)
-\left(\frac{m\lambda}{16\pi^2 L_P}\right)\; K_2^{'}(2L_Pm)\right\}
\end{equation}
and
\begin{equation}
\lambda_{\rm corr}^{\rm P}
=\left(\frac{\partial^4 {\cal V}_{\rm eff}^{\rm P}}{\partial\Phi_c^4}
\right)_{\Phi_c=0}
=\left\{\lambda-\left(\frac{9\lambda^2}{32\pi^2 mL_P}\right)\; 
K_2^{'}(2L_Pm)-\left(\frac{3\lambda^2}{16\pi^2}\right)\; 
K_2^{''}(2L_Pm)\right\},
\end{equation}
where $K_2{'}$ and $K_2^{''}$ denote first and second derivatives of the 
modified Bessel function $K_2$ with respect to the argument, respectively.
In the limit $L_P \to 0$, we obtain,
\begin{eqnarray}
{\left(m_{\rm corr}^{\rm P}\right)\!}^{2} &=&  m^2 +  
\frac{\lambda (2\gamma -1)}{32 \pi^2}m^2 + \frac{\lambda}{32 \pi^2}
\left[\frac{1}{L_P^2} + m^2 \ln(L_P^2m^2) \right] \nonumber \\ 
&& \quad \; \;+ \;\frac{\lambda m^4L_P^2}{128 \pi^2} \left[ \ln(L_P^2m^2) 
+ (4\gamma + 3) \right] \nonumber \\
\lambda_{\rm corr}^{\rm P} &=& \lambda + \frac{3\gamma}{16\pi^2}\lambda^2
+ \frac{3 \lambda^2}{32\pi^2}\ln(L_P^2m^2) \nonumber \\
&& \quad + \; \frac{3\lambda^2 m^2 L_P^2}{32\pi^2} \left[\ln(L_P^2m^2)
+ 2(\gamma -1)\right] \nonumber \\
{\cal V}^P_{\rm finite} &=& \frac{1}{2} m^2\left[\frac{\lambda}{32\pi^2}
\ln\left(1 + \frac{\lambda}{2m^2}\Phi_c^2\right) - \frac{\lambda}{64\pi^2}
\right]\Phi_c^2 \nonumber \\
&& + \; \frac{1}{4!}\left[ \frac{3\lambda^2}{32\pi^2}
\ln\left(1 + \frac{\lambda}{2m^2}\Phi_c^2\right) - 
\frac{9\lambda^2}{64\pi^2} - \frac{11\lambda^2m^2}{192\pi^2}L_P^2 
-\frac{3\lambda^2m^2}{32\pi^2}L_P^2 \ln(L_P^2m^2)\right] \Phi_c^4 
\nonumber \\
&& + \frac{m^4}{64\pi^2}\ln\left(1 + \frac{\lambda}{2m^2}\Phi_c^2\right)
+ \frac{L_P^2}{192\pi^2}\left(m^2 + \frac{1}{2}\lambda \Phi_c^2\right)^3
\ln\left(1 + \frac{\lambda}{2m^2}\Phi_c^2\right) \nonumber \\
&& - \frac{m^6}{192\pi^2}L_P^2 \ln(L_P^2m^2) \label{eqn:lpcorr}
\end{eqnarray}
where $\gamma$ is the Euler-Mascheroni constant.
By comparing Eqn.~(\ref{eqn:corr}) and Eqn.~(\ref{eqn:lpcorr}), it 
is seen that the divergent terms in the expressions for 
$m_{\rm corr}^{\rm P}$ and $\lambda_{\rm corr}^{\rm P}$ are of the 
same form as those for $m_{\rm corr}$ and $\lambda_{\rm corr}$.
There is a quadratically divergent term as well as a 
logarithmic divergence. 
The finite terms appearing in ${\left(m_{\rm corr}^{\rm P}\right)\!}^{2}$
and $\lambda_{\rm corr}^{\rm P}$ which are independent of $L_P$ are different
from that appearing in ${\cal V}_{\rm finite}$ in Eqn.~(\ref{eqn:lpcorr}).
This is because the form of the cutoff used is different.
It is also clear from the above expression that for $m\neq 0$,
${\cal V}^P_{\rm finite}$ is finite in the limit $L_P\to 0$.   
  
\section{Corrections to the effective Lagrangian for 
electromagnetic field}\label{sec:electro}

The system we shall consider in this subsection consists of a 
complex scalar field $\Phi$ interacting with the electromagnetic 
field represented by the vector potential $A^{\mu}$. 
It is described by the following action (see, for e.g., 
Ref.~\cite{ryder85}, p.~98):
\begin{eqnarray} 
{\cal S}[\Phi, A^{\mu}] 
&=& \int d^4x\,{\cal L}(\Phi, A^{\mu})\nonumber\\
&=&\int d^4x\, \biggl\lbrace 
\left(\partial_{\mu}\Phi+iqA_{\mu}\Phi\right) 
\left(\partial^{\mu} \Phi^* - iq A^{\mu}\Phi^*\right)
-m^2\Phi\Phi^*
- {1 \over 4} F^{\mu\nu}F_{\mu\nu}\biggl\rbrace,\label{eqn:emact}
\end{eqnarray}
where $q$ and $m$ are the charge and the mass associated with a 
single quantum of the complex scalar field, the asterisk denotes 
complex conjugation and 
\begin{equation}
F_{\mu\nu} = \partial_{\mu}A_{\nu} - \partial_{\nu} A_{\mu}.
\end{equation}
We shall assume that the electromagnetic field behaves classically, 
hence $A^{\mu}$ is just a $c$-number, while we shall assume the complex 
scalar field to be a quantum field so that $\Phi$ is an operator valued
distribution.
Varying the action~(\ref{eqn:emact}) with respect to the complex scalar 
field $\Phi$, we obtain the following Klein-Gordon equation:
\begin{equation}
\biggl(\left(\partial_{\mu}+iqA_{\mu}\right)
\left(\partial^{\mu}+iqA^{\mu}\right) + m^2\biggl)\Phi=0.\label{eqn:kgem}
\end{equation}
From the above equation it can be easily seen that
\begin{equation}
{\hat H}\equiv\left(\partial_{\mu}+iqA_{\mu}\right)
\left(\partial^{\mu}+iqA^{\mu}\right).
\end{equation}

\subsection{Conventional result}

In what follows we shall evaluate the effective Lagrangian for a constant
electromagnetic background.  
A constant electromagnetic background can be described by the following 
vector potential:
\begin{equation}
A^{\mu}=(-Ez, -By, 0, 0),\label{eqn:vectpot}
\end{equation}
where $E$ and $B$ are constants.
The electric and the magnetic fields that this vector potential gives 
rise to are given by: ${\bf E}=E{\hat z}$ and ${\bf B}=B{\hat z}$,
where ${\hat z}$ is the unit vector along the positive $x$-axis.
The operator ${\hat H}$ corresponding to the vector potential above is 
then given by
\begin{equation}
{\hat H}\equiv\left(\partial_t^2-{\bf \nabla}^2 -2iqEz\partial_t
-2iqBy\partial_x -q^2 E^2+q^2 B^2\right).
\end{equation}
Exploiting the translational invariance of this operator along the 
$t$ and the $x$ coordinates, we can write the corresponding quantum 
mechanical kernel as follows:
\begin{eqnarray}
\lefteqn{K(x, x; s)}\nonumber\\
& &\quad=\;\int_{-\infty}^{\infty}\frac{d\omega}{2\pi}
\int_{-\infty}^{\infty}\frac{dp_x}{2\pi}\nonumber\\
& &\qquad\qquad\quad\times\;
\langle y, z\vert \exp-\left[i\left(-\partial_y^2+(p_x+qBy)^2
-\partial_z^2-(\omega+qEz)^2\right)s\right]\vert y, z \rangle.
\end{eqnarray}
This kernel then corresponds to that of an inverted oscillator along
the $z$-direction (an oscillator with an imaginary frequency) and an
ordinary oscillator along the $y$-direction.
Therefore, using the kernel for a simple harmonic oscillator (see, for
e.g., Ref.~\cite{fandh65}, p.~63) we obtain that
\begin{equation}
K(x, x; s)
=\left\{\left(\frac{1}{16\pi^2 is^2}\right)
\left(\frac{qEs}{{\rm sinh}(qEs)}\right) 
\left(\frac{qBs}{{\rm sin}(qBs)}\right)\right\}.
\end{equation}
Substituting this kernel in the expression for ${\cal L}_{\rm corr}$ in
Eq.~(\ref{eqn:lcorr}), we get
\begin{equation}
{\cal L}_{\rm corr} =  
-\left(\frac{1}{16\pi^2}\right)
\int_{0}^{\infty} \frac{ds}{s^3}\; e^{-i(m^2-i\epsilon)s} 
\left(\frac{qas}{{\rm sinh}(qas)}\right) 
\left(\frac{qbs}{{\rm sin}(qbs)}\right),\label{eqn:lcorrem}
\end{equation}
where $a$  and $b$ are related to the electric and magnetic fields 
${\bf E}$ and ${\bf B}$ by the following relations:
\begin{equation}
a^2-b^2= {\bf E}^2-{\bf B}^2 \qquad{\rm and}\qquad ab={\bf E}\cdot{\bf B}.
\end{equation} 

We can now interpret the real part of ${\cal L}_{\rm corr}$
as the correction to the Lagrangian describing the classical
electromagnetic background given by:
\begin{equation}
{\cal L}_{\rm em} = -\frac{1}{4}F^{\mu\nu}F_{\mu\nu}
=\frac{1}{2} ({\bf E}^2-{\bf B}^2)=\frac{1}{2}(a^2-b^2).
\end{equation}
From Eq.~(\ref{eqn:lcorrem}), it is easy to see that the real part of 
${\cal L}_{\rm corr}$ is given by the expression
\begin{equation}
{\rm Re}~{\cal L}_{\rm corr} 
=  -\left(\frac{1}{16\,\pi^2}\right)
\int_{0}^{\infty} \frac{ds}{s^3}\;\cos(m^2s)\, 
\left({qas \over \sinh(qas)}\right)\;
\left({qbs \over \sin(qbs)}\right).\label{eqn:lcorremreal}
\end{equation}
This quantity can be regularized by subtracting the flat space
contribution which is obtained by setting both $a$ and $b$ in 
the above expression to zero.
Such a regularization then leads us to the following result:
\begin{equation}
{\rm Re}~{\bar {\cal L}}_{\rm corr}
=  -\left(\frac{1}{16 \pi^2}\right)
\int_{0}^{\infty} \frac{ds}{s^3}\;\cos(m^2s)\, 
\left\{\left({qas \over \sinh(qas)}\right)\;
\left({qbs \over \sin(qbs)}\right) - 1\right\}, 
\label{eqn:ralcorre}
\end{equation}
Near $s=0$, the expression in the curly brackets above 
goes as $\left[-q^2s^2(a^2-b^2)/6\right]$.
Hence, ${\rm Re}~{\bar {\cal L}}_{\rm corr}$ is still 
logarithmically divergent near $s=0$. 
But this divergence is proportional to the original 
Lagrangian ${\cal L}_{\rm em}$ and because of this
feature we can absorb this divergence by redefining 
the field strengths and charge. 
Or, in other words, we can renormalize the field strengths 
and charge by absorbing the logarithmic divergence into them
in the following fashion.  
We write
\begin{equation}
{\cal L}_{\rm eff} 
= \left({\cal L}_{\rm em} +  {\rm Re}~{\bar {\cal L}}_{\rm corr}\right) 
=  \left({\cal L}_{\rm em} + {\cal L}_{\rm div}\right) 
+ \left({\rm Re}~{\bar {\cal L}}_{\rm corr} - {\cal L}_{\rm div}\right),
\end{equation}
where we have defined ${\cal L}_{\rm div}$ as follows:
\begin{eqnarray}
{\cal L}_{\rm div} 
&=& -\left(\frac{1}{16 \pi^2}\right)
\int_{0}^{\infty} {ds \over s^3}\, \cos(m^2s) 
\left\{-{1\over 6}q^2s^2\left(a^2-b^2\right)\right\}\nonumber\\
&=& \left(\frac{Z}{2}\right) \left(a^2-b^2\right) 
=\left(\frac{Z}{2}\right)\left({\bf E}^2-{\bf B}^2\right)
=Z\;{\cal L}_{\rm em}
\end{eqnarray}
where $Z$ is a logarithmically divergent quantity described by the integral 
\begin{equation}
Z = \left(\frac{q^2}{48 \pi^2}\right) 
\int_{0}^{\infty} \frac{ds}{s}\; \cos(m^2 s).
\end{equation}
Therefore, we can write
\begin{equation}
{\cal L}_{\rm eff} 
=  \left({\cal L}_{\rm em} + {\cal L}_{\rm div}\right)
+\left({\rm Re}~{\bar {\cal L}}_{\rm corr} - {\cal L}_{\rm div}\right) 
= (1+Z)\, {\cal L}_{\rm em}+{\cal L}_{\rm finite},
\end {equation}
where ${\cal L}_{\rm finite}$ is a finite quantity described by the
integral
\begin{eqnarray}
{\cal L}_{\rm finite} 
&=& \left({\rm Re}~{\bar {\cal L}}_{\rm corr} 
- {\cal L}_{\rm div}\right)\nonumber\\
&=& -\left(\frac{1}{16 \pi^2}\right)
\int_{0}^{\infty}\frac{ds}{s^3}\,\cos(m^2s)\nonumber\\ 
& &\qquad\qquad\qquad\qquad\times\;
\left\{\left(\frac{qas}{\sinh(qas)}\right)
\left(\frac{qbs}{\sin(qbs)}\right) 
- 1 + \frac{1}{6}q^2s^2\left(a^2-b^2\right)\right\}.
\end{eqnarray}  
All the divergences now appear in $Z$.  
Redefining the field strengths and charges as
\begin{equation}
{\bf E}_{\rm phy} = (1 + Z)^{1/2}\, {\bf E}\quad;\quad 
{\bf B}_{\rm phy} = (1 + Z)^{1/2}\, {\bf B}\quad;\quad 
q_{\rm phy} = (1 + Z)^{-1/2}\,q,\label{eqn:physical}
\end{equation}
we find that such a scaling leaves $q_{\rm phy}{\bf E}_{\rm phy} 
=q{\bf E}$ invariant. 
Thus it is possible to redefine (renormalize) the variables in the 
theory thereby taking care of the divergences.  

\subsection{Results with the modified weightage factor}

With the modified weightage factor, we find that the quantity 
${\cal L}_{\rm corr}^{\rm P}$ for the constant electromagnetic
background is described by the following integral:
\begin{equation}
{\cal L}_{\rm corr}^{\rm P}
=  -\left(\frac{1}{16 \pi^2}\right) 
\int_{0}^{\infty} \frac{ds}{s^3}\; e^{-im^2s} \;e^{iL_P^2/s}
\left(\frac{qas}{\sinh(qas)}\right)\;
\left(\frac{qbs}{\sin(qbs)}\right).
\end{equation}
The real part of ${\cal L}_{\rm corr}^{\rm P}$ is then given by
\begin{equation}
{\rm Re}~{\cal L}_{\rm corr}^{\rm P}
=  -\left(\frac{1}{16 \pi^2}\right) 
\int_{0}^{\infty} \frac{ds}{s^3}\; \cos\left[m^2s - (L_P^2/s)\right]\;
\left(\frac{qas}{\sinh(qas)}\right)\;
\left(\frac{qbs}{\sin(qbs)}\right).
\end{equation}
Regularizing this quantity by subtracting the flat space contribution
(viz.~the quantity obtained by setting $a=b=0$ in the above expression), 
we get that 
\begin{equation}
{\rm Re}~{\bar {\cal L}}_{\rm corr}^{\rm P}
=  -\left(\frac{1}{16 \pi^2}\right) 
\int_{0}^{\infty} \frac{ds}{s^3}\; \cos\left[m^2s - (L_P^2/s)\right]\;
\left\{\left(\frac{qas}{\sinh(qas)}\right)\;
\left(\frac{qbs}{\sin(qbs)}\right)-1\right\}.
\end{equation}  
We can now express the effective Lagrangian exactly as we had done 
earlier.
We can define
\begin{equation}
{\cal L}_{\rm eff}^{\rm P}
=\left({\cal L}_{\rm em} 
+ {\rm Re}~{\bar {\cal L}}_{\rm corr}^{\rm P}\right)
=\left({\cal L}_{\rm em} + {\cal L}_{\rm div}^{\rm P}\right)
+\left({\rm Re}~{\bar {\cal L}}_{\rm corr}^{\rm P} 
- {\cal L}_{\rm div}^{\rm P}\right),
\end{equation}
where we can now define ${\cal L}_{\rm div}^{\rm P}$ as follows:
\begin{eqnarray}
{\cal L}_{\rm div}^{\rm P} 
&=& -\left(\frac{1}{16 \pi^2}\right)
\int_{0}^{\infty} \frac{ds}{s^3}\; \cos\left[m^2s - (L_P^2/s)\right]\;
\left\{-\frac{1}{6} q^2 s^2 \left(a^2-b^2\right)\right\}\nonumber\\
&=& \frac{Z^{\rm P}}{2} \left(a^2-b^2\right) 
= \frac{Z^{\rm P}}{2}\,\left({\bf E}^2-{\bf B}^2\right) 
=Z^{\rm P}\, {\cal L}_{\rm em}.
\end{eqnarray}
$Z^{\rm P}$ is now a finite quantity given by the integral
\begin{equation}
Z^{\rm P} = \left(\frac{q^2}{48 \pi^2}\right) 
\int_{0}^{\infty} \frac{ds}{s^3}\; \cos\left(m^2 s-L_P^2/s\right)
=\frac{q^2}{6\pi}\, K_0(2mL_P),
\end{equation} 
where $K_0(2mL_P)$ is the modified Bessel function of order zero.
Therefore, we can write
\begin{eqnarray}
{\cal L}_{\rm eff}^{\rm P}
=\left({\cal L}_{\rm em} + {\cal L}_{\rm div}^{\rm P}\right)
+\left({\rm Re}~{\bar {\cal L}}_{\rm corr}^{\rm P} 
- {\cal L}_{\rm div}^{\rm P}\right)
=\left(1+Z^{\rm P}\right)\, {\cal L}_{\rm em} 
+ {\cal L}_{\rm finite}^{\rm P},
\end{eqnarray}
where ${\cal L}_{\rm finite}^{\rm P}$ is now defined as
\begin{eqnarray}
{\cal L}_{\rm finite}^{\rm P} 
&=& \left({\rm Re}~{\bar {\cal L}}_{\rm corr}^{\rm P} 
- {\cal L}_{\rm div}\right)\nonumber\\
&=& -\left(\frac{1}{16 \pi^2}\right)
\int_{0}^{\infty}\frac{ds}{s^3}\,\cos\left(m^2s-L_P^2/s\right)\nonumber\\
& &\qquad\qquad\qquad\times\; 
\left\{\left(\frac{qas}{\sinh(qas)}\right)
\left(\frac{qbs}{\sin(qbs)}\right) 
- 1 + \frac{1}{6}q^2s^2\left(a^2-b^2\right)\right\}.
\end{eqnarray}  

We can now redefine the field strengths and the charge just as we had
done earlier with $Z^{\rm P}$ instead of $Z$.
$Z^{\rm P}$ is a finite quantity for a non-zero $L_P$, but diverges
logarithmically when $L_P$ is set to zero.
Even in the limit $L_P\to 0$, the quantity ${\cal L}_{\rm finite}^{\rm P}$ 
stays finite and the divergence appears only in the expression for 
$Z^{\rm P}$.

In the limit $L_P\to 0$, we can make a rough estimation of the value of
${\cal L}_{\rm finite}^{\rm P}$ as follows. 
The quantity $\exp(-m^2s -L_P^2/s)$ is a sharply peaked 
function about the value $s=L_P/m \ll 1$.
Therefore, we may, without appreciable error, expand the term in 
the curly brackets in the above expression for 
${\cal L}_{\rm finite}^{\rm P}$ in a Taylor series about the point 
$s=0$ retaining only the first non-zero term.  
Keeping the limits of integration from $0$ to $\infty$, 
(and performing an Euclidean rotation in order to evaluate the 
integral) we obtain,
\begin{equation}
{\cal L}_{\rm finite}^{\rm P} \approx \left(\frac{1}{16\pi^2}\right) 
\frac{2 L_P^2 q^4}{360m^2} \left( 7(a^2-b^2)^2 + 4a^2b^2\right) 
K_2(2mL_P), 
\end{equation}
where $K_2$ is the modified Bessel function of order $2$. 
Expanding $K_2$ in a series and retaining the terms of least order 
in $L_P$, we obtain,
\begin{equation}
{\cal L}_{\rm finite}^{\rm P} \approx \left(\frac{1}{16\pi^2}\right) 
\frac{q^4}{360 m^4} \left( 7(a^2-b^2)^2 + 4a^2b^2\right) 
(1 - L_P^2 m^2).
\end{equation}

\section{Corrections to the thermal effects in the Rindler
frame}\label{sec:rindler}

In flat spacetime, the Minkowski vacuum state is invariant only 
under the Poincare group, which is basically a set of linear 
coordinate transformations.
Under a non-linear coordinate transformation the particle concept,
in general, proves to be coordinate dependent.
For e.g., the quantization in the Minkowski and the Rindler coordinates 
are inequivalent~\cite{fulling73}.
In fact the expectation value of the Rindler number operator in the 
Minkowski vacuum state proves to be a thermal spectrum.
This result is normally obtained in literature by quantizing the field 
in the two coordinate systems and then evaluating the expectation value 
of the Rindler number operator in the Minkowski vacuum state. 
If we are to evaluate the quantum gravitational corrections to the 
Rindler thermal spectrum in such a fashion then we need to know as
to how the metric fluctuations modify the normal modes of the quantum
field.
But, as we have mentioned earlier, we only know as to how quantum 
gravitational corrections can be introduced in the effective Lagrangian.
Therefore, in this section, we shall first evaluate the effective 
Lagrangian in the Rindler coordinates and then go on to evaluate the 
corrections to this effective Lagrangian.
The derivation of the Hawking radiation in a black hole spacetime runs 
along similar lines as the derivation of the Rindler thermal spectrum.
Hence the results we present in this section have some relevance to 
the effects of metric fluctuations on Hawking radiation.
 
The system we shall consider in this section is a massless scalar 
field (in $4$-dimensions described by the action
\begin{equation}
{\cal S}[\Phi] = \int d^4x \sqrt{-g}\;{\cal L}(\Phi)
=\int d^4x\sqrt{-g}\left\{\frac{1}{2}\, g_{\mu\nu}\,
\partial^{\mu}\, \Phi\partial^{\nu}\Phi\right\}.
\end{equation}
Varying this action, we obtain the equation of motion for $\Phi$ to be
\begin{equation}
{\hat H}\Phi\equiv\frac{1}{\sqrt{-g}}
\partial_{\mu}\left(\sqrt{-g}g^{\mu\nu}\partial_{\nu}\right)\Phi=0.
\label{eqn:op}
\end{equation}

\subsection{Conventional result}
 
The transformations that relate the Minkowski coordinates 
in flat spacetime to those of an observer who is accelerating 
uniformly along the $x$-direction are given by the following
relations~\cite{rindler66}:
\begin{equation}
t= g^{-1}(1+g\xi)\, \sinh(g\tau)\;\, ;\;\, 
x= g^{-1}(1+g\xi)\, \cosh(g\tau)\;\, 
;\;\, y=y \;\, ; \;\, z=z,
\end{equation}
where $g$ is a constant.
The new coordinates $(\tau, \xi, y, z)$ are called the
Rindler coordinates.
In terms of the Rindler coordinates the flat spacetime line element
is then given by
\begin{equation}
ds^2=(1+g\xi)^2\, d\tau^2 - d\xi^2 - dy^2 - dz^2.\label{eqn:rindmet}
\end{equation}
Therefore, in the Rindler coordinates, the operator~${\hat H}$ as defined 
in~Eq.~(\ref{eqn:op}) is given by
\begin{equation}
{\hat H}\equiv\left\{\frac{1}{(1+g\xi)^2}\partial_{\tau}^2
-\frac{1}{(1+g\xi)}\partial_{\xi}\left[(1+g\xi)\partial_{\xi}\right]
-\partial_y^2-\partial_z^2\right\},
\end{equation}
where $\partial_x\equiv\left(\partial/\partial x\right)$.
This operator is translationally invariant along the $y$ and the 
$z$-directions. 
Or, in other words, the kernel corresponds to that of a free particle 
along these two directions.
Exploiting this feature, we can write the quantum mechanical kernel as
\begin{equation}
K(x, x, ;s\vert g_{\mu\nu})
=\left(\frac{1}{4\pi is}\right)
\langle \tau, \xi \vert  e^{-i{\hat H'}s}\vert \tau, \xi\rangle,
\end{equation}
where 
\begin{equation}
{\hat H'}\equiv \left(\frac{1}{(1+g\xi)^2}\partial_{\tau}^2
-\frac{1}{(1+g\xi)}\partial_{\xi}
\left[(1+g\xi)\partial_{\xi}\right]\right).
\end{equation}
On rotating the time coordinate $\tau$ to the negative imaginary axis, 
i.e. on setting $\tau=-i\tau_E$, we find that
\begin{equation}
{\hat H'}\equiv
\left(-\frac{1}{(1+g\xi)^2}\partial_{\tau_E}^2
-\frac{1}{(1+g\xi)}\partial_{\xi}
\left[(1+g\xi)\partial_{\xi}\right]\right).
\end{equation}
Now, let $u=g^{-1}(1+g\xi)$, then 
\begin{equation}
{\hat H'}\equiv
\left(-\frac{1}{g^2u^2}\partial_{\tau_E}^2
-\frac{1}{u}\partial_{u}\left[u\partial_{u}\right]\right).
\end{equation}
If we identify $u$ as a radial variable and $g\tau_E$ as an angular
variable then ${\hat H}'$ is similar in form to the Hamiltonian 
operator of a free particle in polar coordinates (in 2-dimensions). 
Then, for a constant $\xi$, the kernel corresponding to the operator 
${\hat H}'$ can be written as~\cite{candd78,paddyBASI}
\begin{equation}
\langle \tau', \xi \vert  e^{-i{\hat H'}s}\vert \tau', \xi\rangle
=\left(\frac{1}{4\pi s}\right)
\sum_{n=-\infty}^{\infty}
\exp \left\{-\frac{i}{4 s}(1+g\xi)^2
\left(\tau-\tau'+2\pi i n g^{-1}\right)^2\right\}.
\end{equation} 

Therefore, the complete quantum mechanical kernel corresponding to 
the operator ${\hat H}$ (in the coincidence limit) is given by
\begin{eqnarray}
K(x, x; s\vert g_{\mu\nu})
&=& \left(\frac{1}{16\pi^2 i s^2}\right)\;
\sum_{n=-\infty}^{\infty}
\exp \left(i\beta^2 n^2/4s\right)\nonumber\\
&=& \left(\frac{1}{16\pi^2 i s^2}\right)\;
\left\{1+ 2 \sum_{n=1}^{\infty} 
\exp \left(i\beta^2 n^2/4s\right)\right\},\label{eqn:kernelrind}
\end{eqnarray}
where $\beta=2\pi g^{-1}(1+g\xi)$.
Substituting this kernel in the expression~(\ref{eqn:lcorr}) and 
setting $m=0$, we find that
\begin{equation}
{\cal L}_{\rm corr}=-\left(\frac{1}{32 \pi^2}\right)\;
\int_{0}^{\infty}\frac{ds}{s^3}\;
\left\{1+ 2 \sum_{n=1}^{\infty} 
\exp \left(i\beta^2 n^2/4s\right)\right\}.
\end{equation}
On regularization, i.e. on subtracting the quantity 
${\cal L}_{\rm corr}^0$ (given by Eq.~(\ref{eqn:lcorr0}) 
with $m=0$) from the above expression, we obtain that
\begin{equation}
{\bar {\cal L}}_{\rm corr}
=\left({\cal L}_{\rm corr}-{\cal L}_{\rm corr}^0\right)
=-\left(\frac{1}{16 \pi^2}\right)\;
\sum_{n=1}^{\infty}\int_{0}^{\infty}\frac{ds}{s^3}\;
\exp \left(i\beta^2 n^2/4s\right).
\end{equation}
The integral over $s$ can be expressed in terms of Gamma functions
(see, for e.g., Ref.~\cite{gandr80}, p.~934), so that
\begin{equation}
{\bar {\cal L}}_{\rm corr}
=\left(\frac{\Gamma(2)}{\pi^2\, \beta^4}\right)
\sum_{n=1}^{\infty} n^{-4}
=\left(\frac{\Gamma(2)}{\pi^2\, \beta^4}\right)\;\zeta(4)
=\left(\frac{\pi^2}{90\, \beta^4}\right),
\end{equation}
where we have made use of the fact that $\zeta(4)=(\pi^4/90)$ 
(cf.~Ref.~\cite{arfken85}, p.~334).

Two points need to be noted regarding the above result.
Firstly, ${\bar {\cal L}}_{\rm corr}$ corresponds to the total 
energy radiated by a black body at a temperature $\beta^{-1}$.
Secondly, there arises no imaginary part to the effective
Lagrangian which clearly implies that the thermal effects 
in the Rindler frame arises due to vacuum polarization and
not due to particle production.
 
\subsubsection{Spectrum from the propagator}

The Feynman propagator corresponding to an operator ${\hat H}$ 
can be written as (cf.~Eq.~(\ref{eqn:prop}))
\begin{equation}
G_{\rm F}(x,x')= -i\int_{0}^{\infty} ds\, 
K(x, x';s\vert g_{\mu\nu}),\label{eqn:gfn}
\end{equation}
where $K(x, x';s\vert g_{\mu\nu})$ is given by Eq.~(\ref{eqn:kernel}).
(Note that since we are considering a massless scalar field
we have set $m=0$.)
For the Rindler coordinates we are considering here, the propagator
is obtained by substituting the kernel~(\ref{eqn:kernelrind}) in
the expression for the propagator as given by~Eq.(\ref{eqn:gfn}).
If we set $\xi=\xi'=0$, $y=y'$ and $z=z'$, we find that the propagator 
is given by  
\begin{eqnarray}
G_{\rm F}(x, x')\equiv G(\Delta\tau)
&=& -\left(\frac{1}{16 \pi^2}\right)
\int_{0}^{\infty} \frac{ds}{s^2} \sum_{n=-\infty}^{n=\infty}
\exp -\left[i(\Delta\tau+i\beta n)^2/4s\right]\nonumber\\
&=& \left(\frac{i}{4\pi^2}\right)
\sum_{n=-\infty}^{n=\infty} \left(\Delta\tau+2\pi i n g^{-1}\right)^{-2},
\end{eqnarray}
where $\Delta\tau=(\tau-\tau')$.
Also, since we have set $\xi=0$, $\beta=(2\pi/g)$. 
(Compare this result with Eq.~(3.66) in Ref.~\cite{bandd82}.)
Fourier transforming this propagator with respect to $\Delta\tau$,
we find that
\begin{eqnarray}
P(\Omega)&\equiv & \Biggl\vert\; \int_{-\infty}^{\infty} d\Delta\tau\, 
e^{-i\Omega \Delta\tau}\, G_{\rm F}(\Delta\tau)\; \Biggl\vert \nonumber\\
&=& \Biggl\vert\; \left(\frac{i}{4\pi^2}\right) 
\int_{-\infty}^{\infty} d\Delta\tau\,
\sum_{n=-\infty}^{n=\infty}
\left(\frac{e^{-i\Omega \Delta\tau}}{(\Delta\tau+i\beta n)^{-2}}\right)\;
\Biggl\vert\nonumber\\
&=&\left(\frac{1}{2\, \pi}\right)
\left(\frac{\Omega}{e^{\beta\Omega}-1}\right),
\end{eqnarray} 
i.e. the resulting power spectrum is a thermal spectrum with a
temperature $\beta^{-1}$.

\subsection{Results with the modified weightage factor}

Let us now evaluate the effective Lagrangian in the Rindler frame with
the modified weightage factor.
Now, substituting the kernel~(\ref{eqn:kernelrind}) in the 
expression~(\ref{eqn:lcorrpl}) for ${\cal L}_{\rm corr}^{\rm P}$ 
and setting $m=0$, we obtain that
\begin{equation}
{\cal L}_{\rm corr}^{\rm P} 
= -\left(\frac{1}{32 \pi^2}\right)\nonumber\\
\int_{0}^{\infty}\frac{ds}{s^3}\; e^{iL_P^2/s}\;
\left\{1+2\sum_{n=1}^{\infty} 
\exp \left(i\beta^2 n^2/4s\right)\right\}.
\end{equation}
On regularization, i.e. on subtracting the quantity 
${\cal L}_{\rm corr}^{{\rm P}0}$ (with $m=0$) from 
${\cal L}_{\rm corr}^{\rm P}$, we find that
\begin{eqnarray}
{\bar {\cal L}}_{\rm corr}^{\rm P} 
&=&\left({\cal L}_{\rm corr}^{\rm P}
-{\cal L}_{\rm corr}^{{\rm P}0}\right)\nonumber\\
&=&-\left(\frac{1}{16 \pi^2}\right)
\int_{0}^{\infty}\frac{ds}{s^3}\; e^{iL_P^2/s}\;
\sum_{n=1}^{\infty} \exp \left(i\beta^2 n^2/4s\right)\nonumber\\
&=&-\left(\frac{1}{16 \pi^2}\right)
\sum_{n=1}^{\infty} \int_{0}^{\infty}\frac{ds}{s^3}\;
\exp \left[i(\beta^2 n^2 + 4L_P^2)/4s\right]\nonumber\\
&=&\left(\frac{1}{\pi^2}\right) 
\sum_{n=1}^{\infty}(\beta^2 n^2 + 4L_P^2)^{-2}\nonumber\\
&=&\left(\frac{1}{\pi^2\beta^4}\right) \sum_{n=1}^{\infty}
\left[n^2 + (4L_P^2/\beta^2)\right]^{-2}.
\end{eqnarray}
Using the relation (\ref{eqn:relationone}),
we can express ${\bar {\cal L}}_{\rm corr}^P$ in a closed form as follows:
\begin{equation}
{\bar {\cal L}}_{\rm corr}^{\rm P}
=\left\{-\left(\frac{1}{32 \pi^2 L_P^4}\right)
+\left(\frac{1}{32 \pi a L_P^3}\right)\; {\rm coth}(2\pi L_P/a) 
+ \left(\frac{1}{16 a^2 L_P^2}\right)\; 
{\rm cosech}^2(2\pi L_P/a)\right\}.
\end{equation}
Making use of the  series expansions (\ref{eqn:relationtwo}) and 
(\ref{eqn:relationthree}), we find that, as $L_P\to 0$
\begin{equation}
{\bar {\cal L}}_{\rm corr}^{\rm P} 
\to \left\{\left(\frac{\pi^2}{90\beta^4}\right) 
- L_P^2\left(\frac{8\pi^4}{945 \beta^6}\right)\right\}.
\end{equation}

\subsubsection{Spectrum from the modified propagator}

The propagator with the modified weightage factor is given by
\begin{equation}
G_{\rm F}^{\rm P}(x,x') 
= -i \int_{0}^{\infty} ds\, e^{i L_P^2/s}\, 
K(x, x';s\vert g_{\mu\nu}).
\end{equation}
In the Rindler coordinates, we find that, if we set $\xi=\xi'=0$, $y=y'$ 
and $z=z'$, the modified propagator is given by  
\begin{eqnarray}
G_{\rm F}^{\rm P}(\Delta\tau)
&=&-\left(\frac{1}{16 \pi^2}\right)
\int_{0}^{\infty} \frac{ds}{s^2} \sum_{n=-infty}^{n=\infty}
\exp -i\left[\frac{1}{4s}(\Delta\tau+i\beta n)^2 
- \frac{L_P^2}{s}\right]\nonumber\\
&=&\left(\frac{i}{4\pi^2}\right)
\sum_{n=-\infty}^{n=\infty} 
\left[(\Delta\tau+2\pi i n g^{-1})^{-2} - 4L_P^2\right]^{-1},
\end{eqnarray}
where, as before, $\Delta \tau =(\tau -\tau')$.
Fourier transforming this modified propagator with respect to 
$\Delta\tau$, we obtain that
\begin{eqnarray}
{\cal P}^{\rm P}(\Omega)
&\equiv& \Biggl\vert\; \int_{-\infty}^{\infty} d\Delta\tau\, 
e^{-i\Omega \Delta\tau}\, 
G_{\rm F}^{\rm P}(\Delta\tau)\;\Biggl\vert \nonumber\\
&=&\Biggl\vert\; \left(\frac{i}{4\pi^2}\right)
\int_{-\infty}^{\infty} d\Delta\tau\, 
\sum_{n=-\infty}^{n=\infty}
\left(\frac{e^{-i\Omega \Delta\tau}}{\left[(\Delta\tau +i\beta n)^2 
- 4L_P^2\right]}\right) \; \Biggl\vert\nonumber\\
&=&\left(\frac{1}{2\, \pi}\right)
\left(\frac{\vert\sin(2\Omega L_P)\vert}{2 \Omega L_P}\right)
\left(\frac{\Omega}{e^{\beta\Omega}-1}\right).
\end{eqnarray} 
This modified shows an appreciable deviation from the Planckian form
only when $\beta \simeq \L_P$.
But when $\beta \simeq L_P$, the semiclassical approximation we are 
working with will anyway cease to be valid.

\section{Corrections to the gravitational Lagrangian}\label{sec:grav}

The system we shall consider in this section consists of scalar field 
$\Phi$ interacting with a classical gravitational field described by the 
metric tensor $g_{\mu \nu}$.
It is described by action
\begin{eqnarray}
{\cal S}[g_{\mu\nu}, \Phi]
&=&\int d^D x\; \sqrt{-g}\, {\cal L}(g_{\mu, \nu}, \Phi)\nonumber\\ 
&=& \int d^Dx\; \sqrt{-g}\,\left\{\frac{1}{16\, \pi G}\, 
(R-2\Lambda)
+ \frac{1}{2} g^{\mu\nu} \partial_{\mu}\Phi \partial_{\nu}\Phi 
- \frac{1}{2}m^2\Phi^2-\frac{1}{2}\xi R\Phi^2\right\},\label{eqn:gravact}
\end{eqnarray}
where $R$ is the scalar curvature of the spacetime, $\Lambda$ is the
cosmological constant and $G$ is the gravitational constant.
Setting the parameter $\xi=0$ or $\xi=(1/6)$
corresponds to a minimal or conformal coupling of the scalar field to 
the gravitational background respectively. 
We are interested in finding quantum corrections to the purely gravitational
part of the total Lagrangian. 
This will be done as usual in the framework of the semiclassical theory 
by considering the one-loop effective action formalism.
In the conventional derivation, divergences arise in the expression 
for ${\cal L}_{\rm corr}$.
There are three divergent terms, two of which are absorbed into the 
cosmological constant $\Lambda$ and the gravitational constant $G$ and 
thus Einstein's theory is reinterpreted suitably.  
The third divergent term cannot be so absorbed. 
Extra terms will have to be introduced into the gravitational Lagrangian
in order to absorb this divergence \cite{bandd82}.
When the duality principle is used however, no divergences occur.
The cosmological constant and the the gravitational constant are 
modified by the addition of finite terms which are seen to diverge
in the limit $\L_P\to 0$ thus recovering the standard result.

\subsection{Conventional result}

In what follows we shall first briefly outline the conventional approach to 
calculating ${\cal L}_{\rm corr}$ and then use the path integral duality
prescription to compute the corrections to the gravitational Lagrangian.

Varying the above action with respect to $\Phi$, we obtain that
\begin{equation}
\left(\frac{1}{\sqrt{-g}}\partial_{\mu}
\left(\sqrt{-g}g^{\mu\nu}\partial_{\nu}\right)
+ m^2 + \xi R\right)\Phi =  0 \label{eqn:fieldeqn}.
\end{equation}
Comparing this equation of motion with Eq.~(\ref{eqn:ofmotion}), it is
easy to identify that
\begin{equation}
{\hat H}\equiv
\left(\frac{1}{\sqrt{-g}}\partial_{\mu} 
\left(\sqrt{-g}g^{\mu\nu}\partial_{\nu}\right) + m^2 + \xi R\right).
\label{eqn:goperator}
\end{equation}

In $D$-dimensions, the quantum mechanical kernel 
$K(x, x'; s\vert g_{\mu\nu})$ (cf.~Eq.~(\ref{eqn:kernel}))
corresponding to the operator ${\hat H}$ above, can be written 
as~\cite{bandd82,parker79}
\begin{equation}
K(x, x'; s\vert g_{\mu\nu})
= \left(\frac{i}{(4 \pi is)^{D/2}}\right)\;  
e^{i\sigma(x,x')/2s}\; \Delta^{1/2}(x,x')\; 
F(x,x';s), \label{eqn:kernelgrav}
\end{equation}
where 
\begin{equation}
\sigma(x,x') 
= \frac{1}{2}\int_0^s ds' \left\{g_{\mu\nu}\frac{dx^{\mu}}{ds'}
\frac{dx^{\nu}}{ds'}\right\}^{1/2}
\end{equation}
is the proper arc length along the geodesic from $x'$ to $x$ and 
$\Delta^{1/2}(x,x')$ is the Van Vleck determinant given by  
\begin{equation}
\Delta^{1/2}(x,x') 
= \Biggl(-[-g(x)]^{-1/2}\; 
{\rm det}[\partial_{\mu}\partial_{\nu}\sigma(x,x')]
\left[-g(x')\right]^{-1/2}\Biggl). 
\end{equation}
The function $F(x,x';s)$ can be written down in an asymptotic expansion
\begin{equation}
F(x,x';s) 
= \sum_{n=0}^{\infty} a_n\, (is)^n 
= a_0 + a_1(x,x')\, (is) + a_2(x,x')\, (is)^2 + \ldots,
\end{equation}
where the leading term $a_0$ is unity since $F$ must reduce 
to unity in flat spacetime. 

Substituting the quantum mechanical kernel above in the expression for
${\cal L}_{\rm corr}$ given by Eq.~(\ref{eqn:lcorr}), we obtain that
\begin{eqnarray}
{\cal L}_{\rm corr} 
&=& -\lim_{x\to x'}\left(\frac{\Delta^{1/2}(x,x')}{32\, \pi^2}\right) 
\int_0^{\infty} \frac{ds}{s^3}\, e^{-i m^2 s}e^{i\sigma(x,x')/2s}\nonumber\\ 
& &\qquad\qquad\qquad\qquad\qquad\quad\times\;
\left\{1 + a_1(x,x')\, (is) + a_2(x,x')\, (is)^2 + \ldots\right\}.
\end{eqnarray}
In the coincidence limit, $\sigma(x,x')$ vanishes and one sees that the 
integral over the first three terms in the square brackets diverge.  
The integral over the remaining terms involving $a_3$, $a_4$ and so on, 
are finite in this limit.  
Therefore, the divergent part of ${\cal L}_{\rm corr}$ is given by
\begin{equation}
{\cal L}_{\rm corr}^{\rm div} 
= -\left(\frac{1}{32\, \pi^2}\right) 
\int_0^{\infty} \frac{ds}{s^3}\, e^{-im^2s}\, 
\left\{1 + a_1(x,x)\, (is) + a_2(x,x)\, (is)^2\right\},
\end{equation}
where the coefficients $a_1$ and $a_2$ are given by the 
relations~\cite{bandd82}
\begin{equation}
a_1(x,x)  =  \left(\frac{1}{6}-\xi \right) R(x)
\end{equation}
and
\begin{equation}
a_2(x,x) 
= \frac{1}{6}\, \left(\frac{1}{5} - \xi\right) 
g^{\mu \nu} R_{;\mu;\nu}(x)  
+  \frac{1}{2}\, \left(\frac{1}{6} - \xi\right)^2\, R^2(x)  
+  \frac{1}{180}\, R_{\mu\nu\lambda\rho} R^{\mu\nu\lambda\rho} 
- \frac{1}{180}\, R_{\mu\nu}R^{\mu\nu}.
\end{equation}

Since $a_1$ and $a_2$ depend only on $R_{\mu\nu\lambda\rho}$ and 
its contractions, they are purely geometrical in nature. 
The divergences arise because of the ultraviolet behavior of the field 
modes.  
These short wavelengths probe only the local geometry in the neighborhood
of $x$ and are not sensitive to the global features of the spacetime and 
are independent of the quantum state of the field $\Phi$. 
Since the divergent part of the effective Lagrangian is purely geometrical 
it can be regarded as the correction to the purely gravitational part of 
the Lagrangian. 
The divergence corresponding to the first term in the square brackets can 
be added to the cosmological constant thus regularizing it while the 
divergence due to the second term can be absorbed into the gravitational 
constant giving rise to the renormalized gravitational constant which is 
finite.  
The third term which involves $a_2$ contains derivatives of the metric 
tensor of order $4$ and this term represents a correction to Einstein's 
theory which contains derivatives of order $2$ only.    
Therefore one needs to introduce  extra terms into the gravitational 
Lagrangian so that these divergences can be absorbed into suitable 
constants~\cite{bandd82}.

\subsection{Results with the modified weightage factor}

Let us now evaluate the corrections to the gravitational Lagrangian with
the modified weightage factor.
Substituting the kernel~(\ref{eqn:kernelgrav}) into 
Eq.~(\ref{eqn:lcorrpl}), we obtain 
\begin{eqnarray}
{\cal L}_{\rm corr}^{\rm P} 
&=&  -\left(\frac{1}{32\, \pi^2}\right) 
\int_0^{\infty} \frac{ds}{s^3}\, e^{-im^2s}\, e^{iL_P^2/ s}\;
\left\{1 + a_1(x,x)\, (is) + a_2(x,x)\, (is)^2 +\ldots\right\}\nonumber\\
&=& \left(\frac{m^4}{32\, \pi)^2}\right) 
\Biggl\{\left(\frac{2}{ L_P^2\, m^2}\right)\, K_2(2L_P m) 
+ \left(\frac{2}{ L_P\, m^3}\right)\, K_1(2L_P m)a_1(x,x)\nonumber\\
& &\qquad\qquad\qquad\qquad\qquad\qquad\qquad\qquad\quad
+\;  \left(\frac{2}{m^4}\right) K_0(2L_P m)a_2(x,x)+\ldots\Biggl\}, 
\end{eqnarray}
where $K_0$, $K_1$ and $K_2$ are the modified Bessel functions of orders 
zero, $1$ and $2$, respectively.   
 
The effective Lagrangian for the classical gravitational background is 
therefore given by
\begin{eqnarray}
{\cal L}_{\rm eff}^{\rm P} 
&=&\left({\cal L}_{\rm grav} +{\cal L}_{\rm corr}^{\rm P}\right)\nonumber\\ 
&=& \frac{1}{16\pi G_{\rm corr}}R - \frac{1}{8\pi G}\Lambda_{\rm corr}
+ \frac{1}{16\pi^2}K_0(2L_P m)a_2(x,x) + \ldots
\end{eqnarray} 
where,
\begin{eqnarray}
\Lambda_{\rm corr} &=& \Lambda - \frac{m^2 G}{2\pi L_P^2} K_2(2L_Pm)
\nonumber \\
\frac{1}{G_{\rm corr}} &=& \frac{1}{G} + \frac{m}{\pi L_P}
\left(\frac{1}{6}-\xi\right) K_1(2L_Pm)
\end{eqnarray}

In Eq.~(\ref{eqn:lcorrpl}), the duality principle demanded that the 
Kernel be modified by  a factor $\exp(iL_P^2/s)$ where $L_P^2$ is the 
square of the Planck length.  
But, as mentioned earlier, $L_P$ could as well be replaced by 
$\eta L_P$ where $\eta$ is a numerical factor of order unity.
Therefore, we replace $L_P$ in the above equations by $\eta L_P$ 
and since $L_P^2 = G$ by definition, the formula for $G_{\rm corr}$ 
can equivalently be written as
\begin{equation}
\frac{1}{G_{\rm corr}} = \frac{1}{G}\left(1+ \frac{m\sqrt{G}}{\eta\pi}
\left(\frac{1}{6}-\xi\right)K_1(2\eta \sqrt{G}m)\right)
\end{equation}

In the limit $(\eta m) \to 0$, using the power series expansion 
for the functions $K_2(2\eta \sqrt{G}m)$ and $K_1(2\eta \sqrt{G}m)$, 
we can write the corrections to $G$ and $\Lambda$ as follows:
\begin{eqnarray}
\Lambda_{\rm corr} 
&=& \Lambda - \frac{m^2}{2\pi \eta^2} 
\left(\frac{1}{2\eta^2 G m^2}-\frac{1}{2}\right) \nonumber \\
&=& \Lambda + \frac{m^2}{4\pi \eta^2} - \frac{1}{4\pi \eta^4 G}
\end{eqnarray}
and
\begin{equation}
\frac{1}{G_{\rm corr}} 
= \frac{1}{G}\left[ 1 + \frac{1}{2\pi \eta^2}\left(\frac{1}{6}-\xi
\right) \right] 
\end{equation}

For the case when the scalar field is assumed to be coupled minimally 
to the gravitational background, i.e. $\xi = 0$, the correction to $G$
reduces to, 
\begin{equation}
\frac{1}{G_{\rm corr}} = \frac{1}{G}\left[1 + 
\frac{1}{12\pi\eta^2}\right]
\end{equation}
while for conformal coupling where $\xi=1/6$, the correction to 
$G$ vanishes.

\section{Corrections to the trace anomaly}\label{sec:trace}

The problem concerning the renormalization of the expectation value of 
energy-momentum tensor in curved spacetime is considerably more 
involved than the corresponding problem in Minkowski spacetime. 
This concerns the role of the energy-momentum tensor $T_{ik}$ in gravity.
In flat spacetime only energy differences are meaningful and therefore 
infinite constants like the energy of the vacuum can be 
subtracted out without any problem.
In curved spacetime, however, energy is a source of gravity.
Therefore, one is not free to rescale the zero point of the energy scale
in an arbitrary manner.
In the semi-classical theory of gravity, one can carry out the 
renormalization of $\langle T_{ik}\rangle$ in a unique way using 
different methods like the $\zeta$-function renormalization technique,
dimensional regularization and other methods. 
Since $\langle T_{ik}\rangle$ can be obtained from the effective action
by functionally differentiating with respect to the metric tensor, 
the renormalization procedure is therefore connected with the 
renormalization of effective action which was described in 
the previous section.
Upon specializing to  theories where the classical action 
is invariant under conformal transformations, it can be shown 
that the trace of the {\it classical} energy-momentum tensor is zero.   
But, when the renormalized expectation value of the trace is calculated, 
however, it is found to be non-zero.  
This is the conformal or trace anomaly.
It essentially arises because of the divergent terms present in 
the effective action. 
When the principle of path integral duality is applied, no 
divergences appear and hence one would expect the trace 
anomaly to vanish.
But because a fixed length scale appears in the problem, the trace anomaly 
is still non-zero.

\subsection{Conventional result}

In this section we derive the formula relating the trace of the energy
momentum tensor and the effective action.  
We then apply the duality principle and derive an explicit formula for 
the trace anomaly.
In the limit of $L_P\to 0$, it is shown that the usual divergences appear
which when renormalized using dimensional regularization and zeta function
regularization techniques yield the usual formula for the trace anomaly.

Consider a scalar field that is coupled to a classical gravitational 
background as described by the action~(\ref{eqn:gravact}).
The energy-momentum of such a scalar field can be obtained by 
varying the action with respect to the metric tensor $g_{\mu\nu}$ 
as follows~\cite{landau2}:
\begin{eqnarray}
T^{\mu\nu} 
&\equiv& \left(\frac{2}{\sqrt{-g}}\right) 
\left({\delta {\cal S} \over \delta g_{\mu\nu}}\right) \nonumber \\ 
&=& \partial^{\mu}\Phi \partial^{\nu} \Phi 
- \frac{1}{2} g^{\mu\nu} \partial^{\alpha} \partial_{\alpha}\Phi 
- \frac{1}{2} g^{\mu\nu} m^2 \Phi^2\nonumber\\
& &\qquad\qquad\qquad\qquad\quad  
+\, \xi \left\{g^{\mu\nu} \frac{1}{\sqrt{-g}}\partial_{\alpha}
\left(\sqrt{-g} g^{\alpha\beta}\partial_{\beta}\right)\Phi^2 
- {\left(\Phi^2\right)}^{;\mu;\nu} + G^{\mu\nu} \Phi ^2\right\},
\end{eqnarray}
where $G^{\mu\nu}=\left(R^{\mu\nu}-(1/2)g^{\mu\nu}R\right)$.  
Using the field equations~(\ref{eqn:fieldeqn}), the trace of 
$T^{\mu\nu}$ is 
\begin{equation}
T^\mu_{\mu} 
= (6\xi -1)\partial^{\mu}\Phi\partial_{\mu}\Phi 
+ \xi(6\xi-1)R \Phi^2 + (6\xi-2) m^2 \Phi^2.
\end{equation}
For the conformally invariant case, i.e. when $\xi=1/6$ and $m=0$, the 
trace vanishes. 
Thus, if the action is invariant under conformal transformations of the 
metric, the classical energy-momentum tensor is traceless. 
Since conformal transformations are essentially a rescaling of lengths 
at each spacetime point $x$, the presence of a mass and therefore the 
existence of a fixed length scale in the theory will break the conformal 
invariance. 

On the other hand, the trace of the renormalized expectation value of 
the energy momentum tensor does not vanish in the conformal limit.  
In the semiclassical case, the expectation value of $T^{\mu\nu}$ is 
given by
\begin{equation}
\langle T^{\mu\nu}\rangle 
= {2 \over \sqrt{-g(x)}}{\delta {\cal S}_{\rm corr}\over 
\delta g_{\mu\nu}},\label{eqn:emtensor}
\end{equation}
where
\begin{equation}
{\cal S}_{\rm corr}=\int d^4x \sqrt{-g} {\cal L}_{\rm corr}.
\end{equation}

Consider the change in ${\cal S}_{\rm corr}$ under an infinitesimal 
conformal transformation
\begin{equation}
g_{\mu\nu}\to \widetilde{g}_{\mu\nu}
= \Omega^2(x)g_{\mu\nu} 
\quad {\rm with}\quad \Omega^2(x)
= 1 + \epsilon(x) \quad {\rm and} 
\quad \delta g_{\mu\nu} 
= \epsilon(x) g_{\mu\nu} \label{eqn:conformal}
\end{equation}
Regarding ${\cal S}_{\rm corr}$ as a functional of 
$\Omega^2(x)$, with $g_{\mu\nu}$ being 
a given function, one obtains,
\begin{equation}
{\cal S}_{\rm corr}[(1 + \epsilon)g_{\mu\nu}] 
= {\cal S}_{\rm corr}[g_{\mu\nu}] 
+ \int \!d^4x \left. {\delta {\cal S}_{\rm corr}[\Omega^2(x) g_{\mu\nu}] 
\over \delta \Omega^2(x)} \right|_{\Omega^2(x)=1} \!\!\!\epsilon(x) 
\end{equation}
Thus,
\begin{equation}
{\delta {\cal S}_{\rm corr}[g_{\mu\nu}] \over \delta g_{\mu\nu}}g_{\mu\nu} 
= \left. {\delta {\cal S}_{\rm corr}[\Omega^2(x) g_{\mu\nu}] \over 
\delta \Omega^2(x)} \right|_{\Omega^2(x)=1} 
\end{equation}
Therefore,
\begin{equation}
\langle T^{\mu}_{\;\; \mu}\rangle 
= {2 \over \sqrt{-g}}{\delta {\cal S}_{\rm corr}[g_{\mu\nu}] 
\over \delta g_{\mu\nu}}g_{\mu\nu}
= {2 \over \sqrt{-g}}\left. {\delta {\cal S}_{\rm corr}
[\Omega^2(x) g_{\mu\nu}] \over \delta 
\Omega^2(x)} \right|_{\Omega^2(x)=1}\label{eqn:trace}
\end{equation} 
Now, ${\cal S}_{\rm corr}$ is given by the  formula 
\begin{equation}
{\cal S}_{\rm corr}[g_{\mu\nu}] 
= -{i\over 2}\int \! d^4x \sqrt{-g} 
\int_0^{\infty} \! {ds \over s} 
\langle x|e^{-is{\hat H}}|x \rangle. \label{eqn:w}
\end{equation}

It can be shown that under a conformal transformation~\cite{bandd82}, 
\begin{equation}
H(x)\to \widetilde{H}(x) = \Omega^{-3}(x) H(x) \Omega(x).
\label{eqn:conformalH} 
\end{equation}
where $H(x)$ is given by Eqn.~(\ref{eqn:goperator}) in the 
conformal limit with $\xi=1/6$ and $m=0$.
The corresponding relation satisfied by the operator ${\hat H}$ under 
a conformal transformation is 
\begin{equation}
\hat{\widetilde{H}} = \Omega^{-1} {\hat H} \Omega^{-1} 
\label{eqn:relation1}
\end{equation} 
while the trace operator $Tr$ defined by the relation 
\begin{equation}
Tr({\hat H}) = \int d^4x \sqrt{-g}\langle x|{\hat H}|x \rangle 
\label{eqn:relation2}
\end{equation}
remains invariant~\cite{parker79}.
Both the above relations can be easily proved as follows.
First note the following definitions. 
In the abstract Hilbert space, one has the relations,
\begin{eqnarray}
\langle x|{\hat H}|x'\rangle &=& H(x)\delta(x-x')[-g(x')]^{-1/2}
\nonumber \\
1 &=& \int d^4x \sqrt{-g}|x\rangle \langle x| \qquad \qquad 
({\rm completeness \; relation}) \nonumber \\
\langle x|x'\rangle &=& [-g(x')]^{-1/2} \delta(x-x') \qquad \;
({\rm Orthonormality \; relation}) \label{eqn:definitions}
\end{eqnarray}
Since the completeness and orthonormality relations remain invariant 
under a conformal transformation, we get, 
\begin{equation}
|x\rangle \to |\widetilde{x}\rangle = \Omega^{-2}(x)|x\rangle .
\end{equation}
Now, using Eqn.~(\ref{eqn:conformalH}) and the definitions given in 
Eqn.~(\ref{eqn:definitions}), 
\begin{eqnarray}
\widetilde{H}(x) \delta(x-x') &=& \Omega^{-3}(x) H(x) 
\Omega(x) \delta(x-x') \nonumber \\
&=& \Omega^{-3}(x)\Omega(x')H(x) \delta(x-x') \nonumber \\
&=& \Omega^{-3}(x) \Omega(x') \sqrt{-g(x')} 
\langle x|{\hat H}|x'\rangle \nonumber \\
&=& \sqrt{-g(x')} \langle x|\Omega^{-3}{\hat H}\Omega |x'\rangle .
\end{eqnarray}
We also have,
\begin{eqnarray}
\widetilde{H}(x) \delta(x-x') &=& \sqrt{-\widetilde{g}(x')} 
\langle \widetilde{x}|\hat{\widetilde{H}}|\widetilde{x}'\rangle
\nonumber \\
&=& \sqrt{-g(x')}\langle x|\Omega^{-2}\hat{\widetilde{H}}\Omega^2|x'\rangle 
\end{eqnarray}
Comparing the above two equations, it is easy to prove the relation 
given in Eqn.~(\ref{eqn:relation1}).
Similarly, using the invariance of the completeness relation under a
conformal transformation, the relation given in 
Eqn.~(\ref{eqn:relation2}) can also be proved. 

Using the above results it is easy to show that  
\begin{equation}
Tr(e^{-is\hat{\widetilde{H}}}) 
= Tr(e^{-is\Omega^{-1}{\hat H}\Omega^{-1}})
= Tr(e^{-is\Omega^{-2}{\hat H}})
\end{equation}
Using the above formula for ${\cal S}_{\rm corr}$, and the above results, 
one finds that under the infinitesimal transformation given 
in Eqn.~(\ref{eqn:conformal}), 
\begin{equation}
{\cal S}_{\rm corr}[\Omega^2 g_{\mu\nu}] 
= -{i\over 2}\int \! d^4x \sqrt{-g} \int_0^{\infty} \! 
{ds \over s} \langle x|e^{-is\Omega^{-2}{\hat H}}|x \rangle
\end{equation}
The above expression for ${\cal S}_{\rm corr}$ is clearly divergent at 
$s=0$ as shown in the previous section. 
Making a change of variable $s\to s'=s\Omega^{-2}$, 
it appears that
\begin{equation}
{\cal S}_{\rm corr}[\Omega^2 g_{\mu\nu}]
= -{i\over 2}\int d^4x \sqrt{-g} \int_0^{\infty} \! 
{ds' \over s'} \langle x|e^{-is'{\hat H}}|x \rangle 
\equiv {\cal S}_{\rm corr}[g_{\mu\nu}]
\end{equation}
But such a change of variable is not valid since the integral 
is divergent.
To make sense of such an integral, one resorts to various 
techniques to determine the trace anomaly. 
As shown in~\cite{bandd82,parker79}, using the $\zeta$ 
function approach, the trace anomaly is shown to be equal 
to $(4 \pi)^{-2} a_2(x,x)$.

\subsection{Results with the modified weightage factor}

But, if one modifies the propagator $\langle x|e^{-is{\hat H}}|x'\rangle$ 
using the duality prescription, then the expression for
${\cal S}_{\rm corr}[\Omega^2 g_{\mu\nu}]$ 
becomes,
\begin{equation}
{\cal S}_{\rm corr}[\Omega^2 g_{\mu\nu}] 
= -{i\over 2}\int d^4x \sqrt{-g} \int_0^{\infty} 
{ds \over s} e^{i L_P^2/s} 
\langle x|e^{-is\Omega^{-2}{\hat H}}|x \rangle \label{eqn:wcp}
\end{equation}
The equation~(\ref{eqn:wcp}) now has no divergences.  
Making a change of variable $s\to s'=s\Omega^{-2}$,which is valid now, 
one obtains,
\begin{equation}
{\cal S}_{\rm corr}[\Omega^2 g_{\mu\nu}] 
= -{i\over 2}\int \! d^4x 
\sqrt{-g} \int_0^{\infty} \! {ds' \over s'}
e^{(i L_P^2/\Omega^2 s')} \langle x|e^{-is'{\hat H}}|x \rangle 
\end{equation}
Using the formula for the trace of the energy momentum 
tensor~(\ref{eqn:trace}), 
one finally obtains 
\begin{equation}
\langle T^{\mu}_{\;\; \mu}\rangle
= -L_P^2 \int_0^{\infty} \! 
{ds \over s^2} e^{iL_P^2/s}\langle x|e^{-is{\hat H}}|x \rangle. 
\end{equation} 
Using the formula for the propagator given in 
Eqn.~(\ref{eqn:kernelgrav}) in $4$ 
dimensions and evaluating the integral over $s$,
\begin{equation}
\langle T^{\mu}_{\;\; \mu}\rangle = {2 L_P^2 \over (4 \pi)^2} 
\sum_{n=0}^{\infty} a_n \left({ L_P \over m} \right)^{n-3} K_{(n-3)}(2 L_P m)
\end{equation}
where $K_n$ is the usual modified Bessel function of order $n$.  
If the limit $L_P \to 0$ is considered, then the expression above 
reduces to 
\begin{equation}
\lim_{L_P \to 0} \langle T^{\mu}_{\;\; \mu}\rangle 
= \lim_{L_P \to 0}{2 \over (4\pi)^2} \left[ {1 \over L_P^4} 
+ { a_1 - m^2 \over 2 L_P^2} \right] 
+  {1 \over 2 (4\pi)^2} (2a_2 - 2m^2a_1 + m^4)
\end{equation}
The terms present in the square brackets represent the divergences that 
are present in the evaluation of the energy-momentum tensor without using 
the duality principle.  
These divergences need to be regularized by other methods like the 
$\zeta$ function approach mentioned earlier.  
The finite part that remains is the last term that, in the conformal 
limit, reduces to $(4 \pi)^{-2} a_2(x,x)$. 
Therefore, in the limit of $L_P \to 0$, the standard result is recovered.  
  
\section{Discussion}\label{sec:cnclsns}

In this paper, we had evaluated the quantum gravitational corrections to 
some of the standard quantum field theoretic results using the `principle 
of path integral duality'.
We find that, the main feature of this duality principle, as we had 
stressed earlier in the introductory section, is that it is able provide 
an ultra-violet cut-off at the Planck energy scales thereby rendering the 
theory finite.
One key feature of thisapproach is that the prescription is completely
Lorentz invariant. Hence we could obtain finite but Lorentz invariant results
for otherwise divergent expressions.

The obvious drawbacks of the approach are the following: 

(i) The prescription 
of path integral duality is essentially an ad-hoc prescription.
It is not backed by a theoretical framework which is capable of replacing
the conventional quantum field theory at the present juncture. 
Hence, the prescription only tells us how to modify the kernels and 
Green's functions. 
To obtain any result with the prescription of path integral duality we
have to first relate the result to the kernel or Green's function,
modify the kernel and thereby obtain the final result.

In spite of this constraint we have been able to show in this paper
that concrete computations can be done and specific results can be 
obtained.
As regards the ad-hocness of the prescription, it should be viewed
as the first step in the approach to quantum gravity based on a 
general physical principle.
It's relation with zero-point length and the emergence of analogous
duality principles in string theories, for example, makes one hopeful
that it can be eventually put on firmer foundation.

(ii) The modified kernel (based on the principle of path integral 
duality) may not be obtainable from the standard framework of field
theory based on unitarity, microscopic causality and locality.
(We have no rigorous proof that this is the case; however, it is quite
possible since standard field theories based on the above principles
are usually divergent.)

It is not clear to us whether such principles will be respected in the 
fully quantum gravitational regime.  
It is very likely that the continuum field theory which we are 
accustomed to will be drastically modified at Planck scales.
If that is the case, it is quite conceivable that the quantum
gravitational corrections also leave a trace of the breakdown of 
continuum field theory even when expressed in such a familiar 
language.
As an example, consider an attempt to study and interpret quantum
mechanics in terms of classical trajectories.
Any formulation will lead to some contradictions like, for example,
the breakdown of differentiability for the path.
This arises because we are attempting to interpret physical principles
using an inadequate formalism.

The next logical step will be to attempt to derive path integral duality
from a deeper physical principle using appropriate mathematical methods.
This should throw more light on, for example, the connection between 
path integral duality and zero-point length which at the moment remains
a mystery. 
We hope to address it in a future publication. 

\section*{\centerline {Acknowledgments}}

\noindent
KS is being supported by the Senior Research Fellowship of the 
Council of Scientific and Industrial Research, India.

\end{document}